\documentclass[a4paper,aps,prx,floatfix,twocolumn, superscriptaddress,showpacs,nobibnotes,longbibliography,nofootinbib]{revtex4-2}

\usepackage[pdftex]{graphicx}
\setkeys{Gin}{width=0.9\columnwidth}

\usepackage{times}
\usepackage{latexsym}
\usepackage{verbatim}
\usepackage{bbold}
\usepackage{bbm}
\usepackage{lipsum}

\usepackage{amsmath}
\usepackage{amsfonts}
\usepackage{color}
\usepackage[english]{babel}
\usepackage{microtype}
\usepackage{comment}
\usepackage[normalem]{ulem}
\usepackage{float}
\usepackage{derivative}
\usepackage{tabularx}
\usepackage{gensymb}
\newcolumntype{Y}{>{\centering\arraybackslash}X}
\usepackage{fancyhdr}
\usepackage[caption=false, position=top, font={bf,footnotesize}, justification=raggedright, singlelinecheck=false]{subfig}

\usepackage{soul,xcolor}
\setstcolor{red}

\usepackage{relsize}

\usepackage[hidelinks,unicode=true]{hyperref}
\hypersetup{
colorlinks=true,
linkcolor=blue,
urlcolor=blue,
citecolor=blue,
pdfhighlight=/N}

\usepackage{comment}

\renewcommand{\sout}[1]{\unskip}

\usepackage{relsize}
\newcommand{\MS}[1]{{\color{cyan} (MS: {#1})}}
\newcommand{\NoteJ}[1]{{\color{magenta} (JO: {#1})}}

\begin{document}

\title{Charting multidimensional ideological polarization across demographic groups in the United States}


\author{Jaume Ojer}
\affiliation{Departament de F\'isica, Universitat Polit\`ecnica de Catalunya, Campus Nord, 08034 Barcelona, Spain}

\author{David C\'arcamo}
\affiliation{Departament de F\'isica, Universitat Polit\`ecnica de Catalunya, Campus Nord, 08034 Barcelona, Spain}

\author{Romualdo Pastor-Satorras}
\affiliation{Departament de F\'isica, Universitat Polit\`ecnica de Catalunya, Campus Nord, 08034 Barcelona, Spain}

\author{Michele Starnini}
\affiliation{Department of Engineering, Universitat Pompeu Fabra, 08018 Barcelona, Spain}
\affiliation{CENTAI Institute, 10138 Turin, Italy}
\email{Corresponding author: michele.starnini@upf.edu}


\date{\today}

\begin{abstract}
Has ideological polarization actually increased in the last decades, or have voters simply sorted themselves into parties matching their ideology more closely? 
We present a novel methodology to quantify multidimensional ideological polarization, by embedding the respondents to a wide variety of political, social, and economic topics from the American National Election Studies (ANES) into a two-dimensional ideological space. 
By identifying several demographic attributes of the ANES respondents, 
{we chart how political and socio-economic groups move through the ideological space in time.} 
{We observe that income and especially racial groups align into parties,
but their ideological distance has not increased over time. 
Instead, Democrats and Republicans have become ideologically
more distant in the last 30 years: Both parties moved away from the center, at different rates. 
Furthermore, Democratic voters have become ideologically more heterogeneous after 2010, indicating that partisan sorting has declined in the last decade.} 
\end{abstract}

\maketitle



The heated debate among political scientists over whether ideological polarization has intensified in America is still ongoing~\cite{abramowitz_is_2008, fiorina_polarization_2008, abramowitz_polarized_2013}. 
Ideological polarization refers to 
individuals having divergent beliefs on ideological issues, such as abortion or affirmative action~\cite{dimaggio_have_1996}. 
Some works conclude that Americans have moved toward the most extreme ideological positions of the political spectrum in recent decades~\cite{abramowitz_disappearing_2010, campbell_polarized_2016}. 
Social media have been pointed out as an underlying root for this increasing disagreement~\cite{bail_exposure_2018, van_bavel_how_2021}. 
Recommendation algorithms, in particular, are suspected to reinforce ideological segregation~\cite{santos_link_2021}, reducing engagement with information from opposing viewpoints, a phenomenon known as echo-chambers~\cite{flaxman_filter_2016, baumann19, cinelli_echo_2021, diaz-diaz_echo_2022}. 
Very recently, however, this hypothesis has been disputed, {after observing that} changing Facebook's feed algorithm to reduce exposure to like-minded content does not seem to reduce the political polarization of users~\cite{garcia-nature, nyhan_like-minded_2023, guess_how_2023}.

Other researchers, instead, argue that the perception of widespread ideological polarization in American politics is exaggerated and primarily driven by the behavior of political elites and the media~\cite{fiorina2005culture}. 
They point to partisan-ideological sorting as a cause for the increasing divide between Democrats and Republicans, by which people sort into the ``correct" combination of party and ideology~\cite{levendusky_partisan_2009}. 
Over the past forty years, there has been a substantial increase in the relationship between party identification and ideological and social identification~\cite{huddy_expressive_2015, mason_uncivil_2018, west_partisanship_2022}, and in the relationship between party identification and positions on a wide range of specific policy issues~\cite{fiorina_political_2008}. 
{Partisan sorting contributes to shaping ideological consistency of the electorate, together with the correlations between their attitudes toward political issues~\cite{converse_nature_2006}.}

{In the field of social psychology, researchers have argued that partisan sorting has been 
responsible for increased levels of partisanship and polarized behavior, including partisan bias, activism, and anger~\cite{mason_i_2015}. 
They claim that partisan-ideological sorting has 
increased affective polarization in the United States~\cite{mason_cross-cutting_2016} and other countries~\cite{harteveld_ticking_2021}, while ideological polarization has not changed much.} 
Within this framework, groups with strong social identities, such as racial, religious, or ideological identities, firmly align with parties in recent decades~\cite{egan_identity_2020}. 
For instance, the average correlations between different issue attitudes and party identification (Democratic or Republican) between 1972 and 2004 increased by 5 percentage points per decade~\cite{baldassarri_partisans_2008}. 
Other studies found correlations with race and religion, being black and atheist Americans aligned with the Democratic Party, while white and Christian groups with the Republican Party~\cite{mason_one_2018}. 
Likewise, Americans with higher incomes tend to hold more conservative preferences on economic policy, but more liberal stances on social policy, and vice versa~\cite{wright_income_2020}. 
Interestingly, online interactions on social media are segregated by demographic factors~\cite{monti_evidence_2023}, and contribute to the rise of affective polarization through partisan sorting~\cite{tornberg_how_2022}.

To discern these underlying factors, researchers from different disciplines, ranging from political and social science to physics and computer science, {have recently} embarked on the endeavor of quantifying ideological polarization~\cite{esteban_measurement_1994, boxell_greater_2017, hohmann_quantifying_2023}. 
On the one hand, they leverage data from social media to define new methodologies to measure the extent of polarization~\cite{guerra_measure_2013, morales_measuring_2015, garimella_quantifying_2018, waller_quantifying_2021}. 
However, social media users are usually not representative of the whole population~\cite{zagheni_demographic_2015}. 
For instance, Facebook and Twitter users are demographically biased, being younger, 
more liberal, and better educated {on average}~\cite{mellon_twitter_2017}. 
On the other hand, several measures of ideological polarization and partisan sorting have been proposed on offline data~\cite{miller_red_2015, davis_party_2016, bougher_correlates_2017}. 
However, these methods are mainly based on 
{self-reported scores with respect to the degree of
partisanship or the ideological identity strength,} 
often including substantial measurement error due to social desirability bias~\cite{brenner_lies_2016}. 
Similarly, partisanship measured by self-reported scores might be biased, even by the structure of the questionnaire itself~\cite{schiff_priming_2022}. 
Furthermore, most of these works encompass single topics, while ideological polarization 
{embraces multiple issues at the same time~\cite{klar_multidimensional_2014}, thus requiring a multidimensional modelling framework~\cite{baumann_emergence_2021, pedraza_analytical_2021}.} 
When dealing with several topics simultaneously, the discrete nature of the Likert scale {---commonly used to assess opinions in research questionnaires---}
makes it even harder to interpolate ideological polarization.

In this paper, we propose a novel methodology to 
quantify partisan sorting and the evolution of ideological polarization,  
{simultaneously with respect to different topics.}
{To this aim, we focus on the respondents of the American National Election Studies (ANES)~\cite{anes}, encompassing opinions regarding a large number of political, social, and economic issues. 
We take into account the political leaning (Democratic or Republican parties) and five demographic attributes (Race, Gender, Age, Affluence, and Education) of the ANES respondents, 
and embed their opinions into a two-dimensional ideological space.} 
There, we can gauge {ideological} polarization by simply measuring the Euclidean distance between opposite political and demographic groups. 
{We can measure partisan sorting by the heterogeneity of the opinion distribution in the ideological space: Ideologically more consistent parties will occupy smaller regions~\cite{garner_polarization_2011}, 
a sign of high ideological coherence in matching political leaning with issue preferences~\cite{baldassarri_partisans_2008}.} 
{Furthermore, we chart the identification of socio-economic groups into parties in the ideological space, depicted as the alignment 
between parties and different demographic groups.}

Our results 
{show that racial and income groups are aligned into parties, but the ideological distance between these demographic groups did not increase in the last 30 years.
Instead, we observe that Democrats and Republicans have become ideologically more distant over time. 
Furthermore, Democratic voters became more heterogeneous after 2010: A part {occupied} a novel region of the ideological space, connected with opinions regarding minority rights. 
These observations directly contradict the hypothesis of sorted, more homogeneous parties in the last decade.}


\section*{Results}

We focus on the ANES surveys conducted in the years 1992, 2000, 2008, 2016, and 2020. 
The last five United States Presidents were elected these years: Clinton (Democratic), Bush (Republican), Obama (D), Trump (R), Biden (D). 
We preprocess data to select questions related to different topics such as the state of the economy, government spending, religion, or minority rights. 
{By picking common questions across years, we can study the temporal evolution of the respondents' opinions.} 
{We refrain from selectively choosing questions explicitly associated with ideological polarization, instead, we include all questions that meet quantitative criteria concerning scales and missing values,} 
see ``Methods'' section ``The ANES dataset'' for details. 
Our working dataset consists of $N=11614$ respondents answering 29 different questions. 
{This dataset includes questions related to policy (e.g., ``Should federal spending on aid to poor people be increased, decreased, or kept about the same?"), social identity (``Which of these statements comes closest to describing your feelings
about the Bible?"), and personal beliefs or attitudes of the respondents (``Would this country be better off if we worried less about how equal people are?"). 
The questions explore the respondents' values regarding government intervention, social equality, racial justice, and personal responsibility. 
In this way, we aim to build a broader ideological profile based not only on direct policy preferences, but also on cultural beliefs and moral values that might indirectly influence opinions concerning political or governmental actions.  
While we believe that this enriched representation can better capture the ideological spectrum of the respondents, we also check that our results are consistent by selecting only a subset of questions specifically related to policy.}

{To} compare the opinions of different groups of individuals, 
we leverage the rich metadata in the ANES survey to identify five demographic attributes (Race, Gender, Age, Affluence, and Education) and the political leaning 
of the respondents. 
We select two opposite groups per attribute, reported in Table~\ref{tab:attributes}, in order to highlight the differences in their opinions. 
For instance, we consider Low-Income and High-Income groups if the respondent belongs to a percentile lower than 34\% or higher than 66\%, respectively, of the American income distribution. 
Individuals in the middle-income group are excluded. 
With respect to the political leaning, we consider Democrats and Republicans (as per self-attribution in ANES data) excluding individuals classified as Independents. 
Fig.~3 in the Supplementary Material (SM) shows the proportion of each group for the different attributes, over the years. 

{In the following, we will first describe the opinion embedding into the ideological space and show how parties and demographic groups occupy different regions. 
We can thus quantify the degree of identification of demographic groups into parties, and the ideological distances between them. 
Furthermore, we will study the evolution of trajectories over time, showing how groups move in the ideological space over the years. 
Finally, we will show how partisan sorting decreased in the last decade, as Democratic voters became less ideologically homogeneous, occupying a novel region of the ideological space.}

\begin{table}[tbp]
\begin{tabular}{p{0.29\columnwidth}|p{0.28\columnwidth}p{0.28\columnwidth}}
\hline
\hfil Attribute & \multicolumn{2}{c}{Groups} \\
\hline
\hfil Party & \hfil Democrats & \hfil Republicans \\
\hfil Race & \hfil Black & \hfil White \\
\hfil Gender & \hfil Female & \hfil Male \\
\hfil Age & \hfil 17-34 & \hfil 55+ \\
\hfil Affluence & \hfil Low-Income & \hfil High-Income \\
\hfil Education & \hfil No-College & \hfil College \\
\hline
\end{tabular}
\caption{Classification of the respondents by attributes according to two opposite groups.}
\label{tab:attributes}
\end{table}

\subsection*{Embedding opinions into an ideological space}

{A possible} 
data representation is a multidimensional space containing 11614 points (respondents) in 29 dimensions (questions). 
Each data point corresponds to a vector in the space of the 29 questions selected (forming the axes), representing the opinion of a single respondent. 
However, this high-dimensional representation is sparse: 
As the number of dimensions increases, the volume of available space grows exponentially and every data point becomes increasingly separate from the rest~\cite{aggarwal_surprising_2001}. 
As a consequence, distance metrics like Euclidean distance lose their usefulness in high-dimensional spaces~\cite{altman_curses_2018}.

Therefore, we reduce the dimensionality of the data representation 
by embedding the opinions of ANES respondents into a two-dimensional \emph{ideological space}. 
The opinion embedding allows us to chart and quantify ideological polarization by using a simple Euclidean distance. 
Furthermore, it allows us to interpolate the discrete nature of the original data, where opinions are expressed in the Likert scale, into a continuous ideological space. 
{In the last years, researchers have applied different machine learning techniques to estimate the structure of preferences within American political institutions~\cite{grimmer_machine_2021}.} 
Similar dimensionality reduction methods have been used to map the opinion similarity and ideological leaning of Internet users~\cite{faridani_opinion_2010, monti_learning_2021}, as well as the scientific knowledge landscape into a low-dimensional spatial representation~\cite{chinazzi_mapping_2019, singh_charting_2024}. 
By studying the embedding trajectories of users in this space one can, for instance, predict their future evolution in time~\cite{kumar_predicting_2019}.


In this work, we implement Isomap~\cite{tenenbaum_global_2000}, a non-linear, quasi-isometric, low-dimensional embedding algorithm 
(see ``Methods'' section ``Dimensionality reduction'' for more details), to obtain a two-dimensional data representation. 
{Unlike other embedding methods, Isomap is designed to preserve large-scale structures of the original data~\cite{saul_manifolds_2003}, making it valuable for examining the global properties of the ideological space~\cite{tenenbaum_global_2000, cox_multidimensional_2000}.} 
Since we select the same set of 29 questions for each year, we embed all opinions across years into the same ideological space, 
{which allows us} 
to study the temporal evolution of opinions. 
{Note that the spatial orientation of the ideological space is arbitrary, 
with the two axes representing a (non-linear) combination of the 29 questions selected.} 
{In this setting, the Euclidean distance between two points is the only meaningful information,  measuring the ideological distance between the corresponding individuals: The larger the distance, the less akin their opinions. 
Fig.~4 in the SM shows the density of respondents in the ideological space. 
Within this framework, partisan sorting can be quantified by the heterogeneity of the opinion distribution: More ideologically consistent and homogeneous parties will occupy smaller regions. 
Furthermore, we can measure the alignment of the opinions of different political and demographic groups: Groups with similar stances on multiple topics will occupy the same regions in the ideological space.}

{As a first step,} 
we checked that the opinion embedding does not lose part of the information contained in the original data, by comparing the accuracy of a classification algorithm on original and embedded data.
We predict the political leaning of respondents (Republican or Democratic) by using logistic regression informed by i) their responses to the full set of 29 questions, and ii) their coordinates in the ideological space.
The ten-fold cross-validated balanced accuracy reads $0.76 \pm 0.06$ for original data, and $0.75 \pm 0.06$ for embedded data (see ``Methods'' section ``Dimensionality reduction'' for details). 
Furthermore, we tested that our findings (shown in the following) do not depend on the specific embedding representation, by using a different value of the Isomap hyperparameter $K$ and by implementing other embedding algorithms, 
{namely PCA~\cite{jolliffe_principal_2002}, t-SNE~\cite{vanderMaaten_visualizing_2008}, and UMAP~\cite{mcinnes2020umap}}, see SM.
{Finally, we also checked that our results hold by embedding opinions into a higher dimensional ideological space, formed, e.g., by three or four dimensions, 
see SM.
The cross-validated accuracy in predicting the political leaning of respondents in three- and four-dimensional embeddings is the same as in the two-dimensional case.
}

\subsection*{Ideological distance between demographic groups}

\begin{figure}[tbp]
    \centering
    \includegraphics[width=0.95\columnwidth]{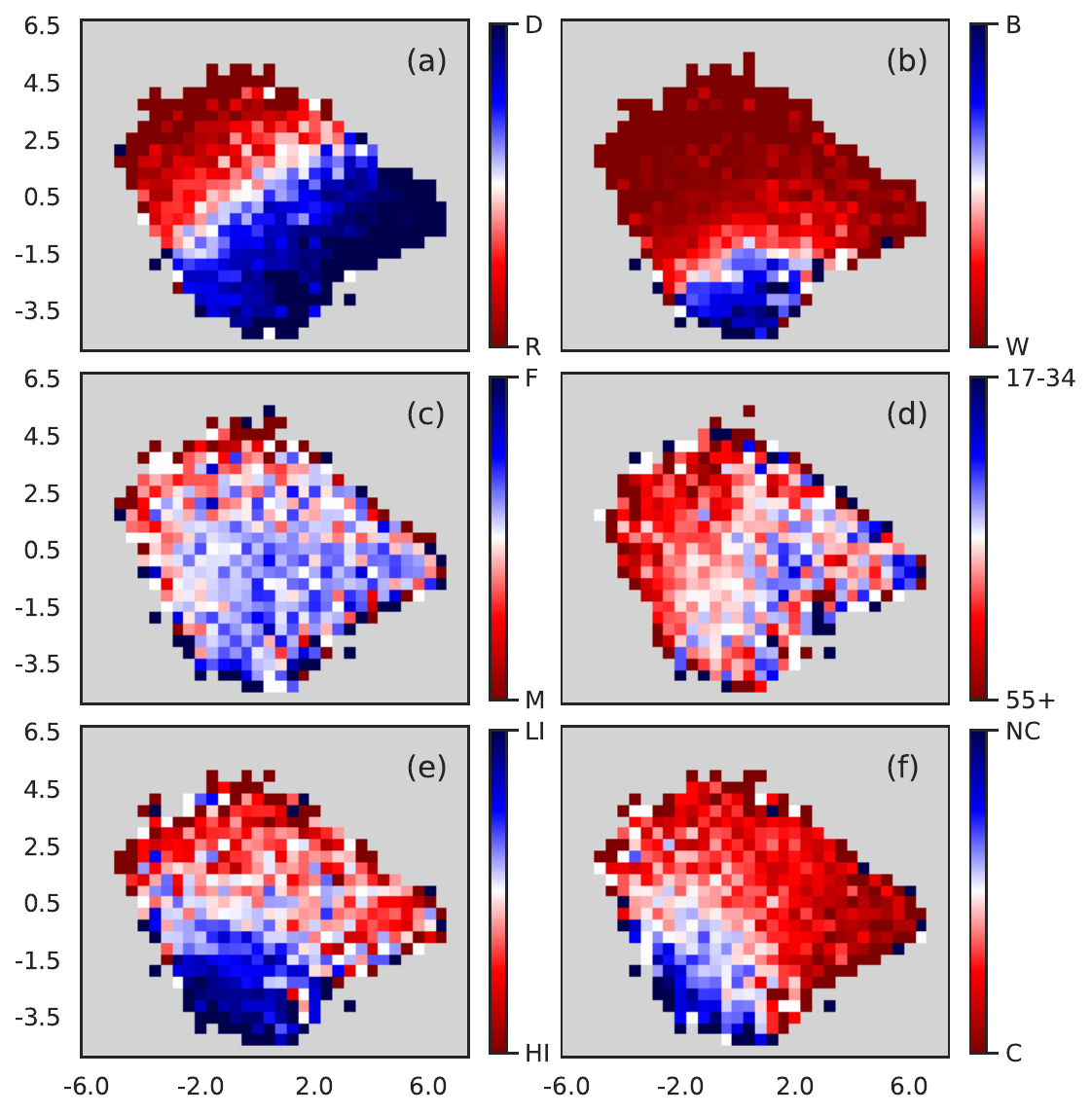}
    \caption{Density distribution of opposite
    groups in the ideological space, for different attributes: (a) Democrats (D) and Republicans (R), (b) Black (B) and White (W), (c) Female (F) and Male (M), (d) aged 17-34 and 55+, (e) Low-Income (LI) and High-Income (HI), (f) No-College degree (NC) and College (C). 
    Darker (lighter) cells have a strong (weak) prevalence of one group with respect to the other. 
    Cells with no respondents are colored in gray.
    We bin the ideological space into 34 $\times$ 29 cells. 
    }
    \label{fig:respondents_density}
\end{figure}

Fig.~\ref{fig:respondents_density} shows the density distribution of opposite groups in the ideological space, for different attributes. 
For each attribute, we assign a binary 
variable $\{+1, -1 \}$ to opposite groups, e.g., 
a value $+1$ to Republicans and $-1$ to Democrats. 
We then divide the ideological space in a lattice of small cells and compute the average of the binary variable in each cell. 
Cells with a large majority of one group are colored by darker colors, while cells in lighter colors are populated by roughly the same number of individuals of the two groups. 
For instance, in Fig.~\ref{fig:respondents_density}(a), dark red (blue) cells are mostly populated by Republicans (Democrats).

One can see that the opinions of different groups are not randomly distributed in the space, but rather ordered for most attributes: 
{One group (for instance, Democrats) tends to occupy a well-defined region of the ideological space, while the opposite group (Republicans) is more present in the opposite region.} 
The attributes of Party, Race, Affluence, and Education show the most ordered distributions, with opposite groups clearly separated in the ideological space, while 
the opinion distributions with respect to Gender and Age are much more uniform. 
This indicates that Republicans and Democrats, for instance, have different and often opposite opinions with respect to the 29 questions selected. 
While this observation can be checked at the level of the single question, our method allows us a bird-eye view of ideological polarization with respect to many different topics at the same time.

Furthermore, not only opposite groups are segregated in the ideological space, but they are also ordered along a certain direction. 
The Black, Low-Income, and No-College groups mostly populate the bottom region of the embedding, whereas the White, High-Income, and College groups are mostly present at the top of it. 
This observation can be quantified by computing the gradient of the different opinion distributions throughout the ideological space. 
For each cell of Fig.~\ref{fig:respondents_density}, we compute the direction and rate of the fastest increase of the average opinion. 
We use finite differences to approximate the derivatives of the gradient with high accuracy\footnote{We use the \textit{gradient} function from the open-source Python library \textit{numpy}~\cite{numpy2020array}.}. 
We define the gradient of the opinion distribution as the vector sum of the gradient values of each cell. 
A longer (shorter) gradient vector is obtained, thus, if the opinion distribution is more (less) ordered along a certain direction. 

\begin{figure}[tbp]
    \centering
    \includegraphics[width=0.8\columnwidth]{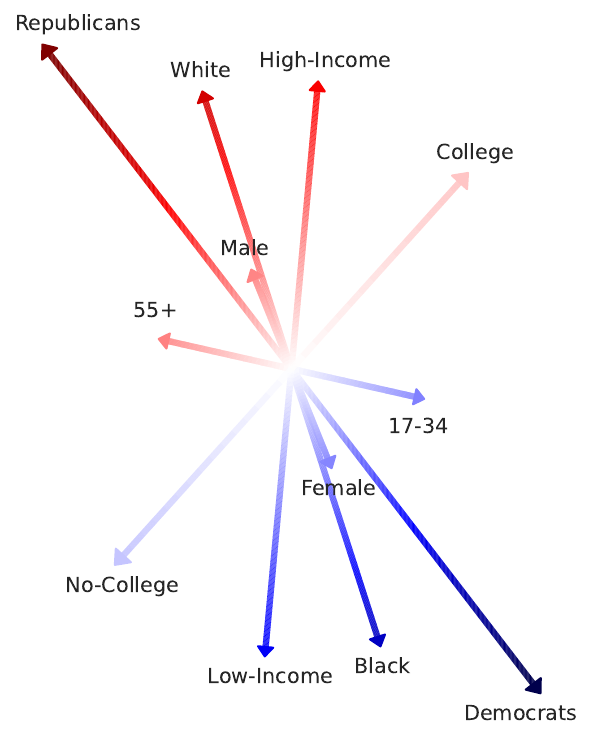}
    \caption{Gradient vectors for each attribute computed from their opinion distribution shown in Fig.~\ref{fig:respondents_density}.
    We discard cells with less than 10 respondents.
    We slightly smooth data by applying a different Gaussian filter with a low standard deviation for each attribute.
    The blue-red colormap obtained for the Party attribute is used, with Republicans in dark red and Democrats in dark blue. 
    The vectors are colored according to the Republicans-Democrats axis, i.e., the color of each point in the vectors is proportional to the radius times the cosine of their angle with the Republicans-Democrats axis.
    }
    \label{fig:respondents_density_arrows}
\end{figure}

Fig.~\ref{fig:respondents_density_arrows} shows the resulting gradients for each attribute, indicating that groups whose opinions are more polarized are Democrats vs Republicans, Low-Income vs High-Income, Black vs White, and No-College vs College, in this order. 
The attributes of Gender and Age, instead, show little polarization, as indicated by short gradient vectors. 
In Fig.~\ref{fig:respondents_density_arrows}, we color each vector by the Party colormap, with Republicans in dark red and Democrats in dark blue. 
The remaining vectors are colored according to the Republicans-Democrats axis.
As a consequence, the Education (College to No-College) vector, almost orthogonal to the Party one, is very lightly colored.

The gradient vectors are useful to quantify 
{the identification of different demographic groups into parties. 
Indeed, one can compute the degree of partisan alignment by means of the cosine similarity, $S_c$, between vectors.} 
Party and Race gradient vectors point in similar directions {with $S_c = 0.94$}, indicating that the opinions of Black (White) individuals significantly overlap with those of Democrats (Republicans). 
The same observation partially holds for the Affluence attribute {($S_c = 0.72$)}, opinions of High-Income (Low-Income) individuals are more similar to Republicans (Democrats).
Instead, the Education vector is almost orthogonal to the Party one {($S_c = 0.17$)}, indicating little 
{correlation between the educational scale and partisan leaning. 
We remark that this observation is valid over time-aggregated data, i.e., over a temporal horizon of 30 years. 
Recently, instead, well-educated individuals increasingly identified with or leaned toward the Democratic Party~\cite{pew_deep_2015}.} 
Therefore, one could identify the Party-Education attributes as the two main axes of the ideological space, 
and express the likelihood of individuals to belong to the four groups defined by these axes. 
Individuals in the upper (lower) region of the ideological space are more likely to be Republicans (Democrats) with (without) a college degree. 
Looking at Fig.~\ref{fig:respondents_density}(b) and (e), one can see that this upper (lower) region is mostly populated by White (Black) and High-Income (Low-Income) individuals.  

The correlation between the opinions of different demographic groups observed so far can be quantified more precisely. 
{In Fig.~5 in the SM we compare 
the opinion distributions in the ideological space of the most polarized attributes (i.e., Party, Affluence, Race, Education) by means of the two-dimensional Kolmogorov-Smirnov (KS) test~\cite{peacock_two-dimensional_1983}. 
In this way, we identify the demographic groups with similar opinion distribution, i.e., showing a certain degree of ideological affinity~\cite{bush_politics_2017, martin-gutierrez_multipolar_2023}. 
Fig.~6 in the SM shows 
that different socio-economic groups can be ordered in terms of ideological affinity, from Republican, White, and High-Income groups to Low-Income, No-College, and Black groups.}

\subsection*{Charting ideological trajectories in time}

\begin{figure}[tbp]
    \centering
    \includegraphics[width=0.95\columnwidth]{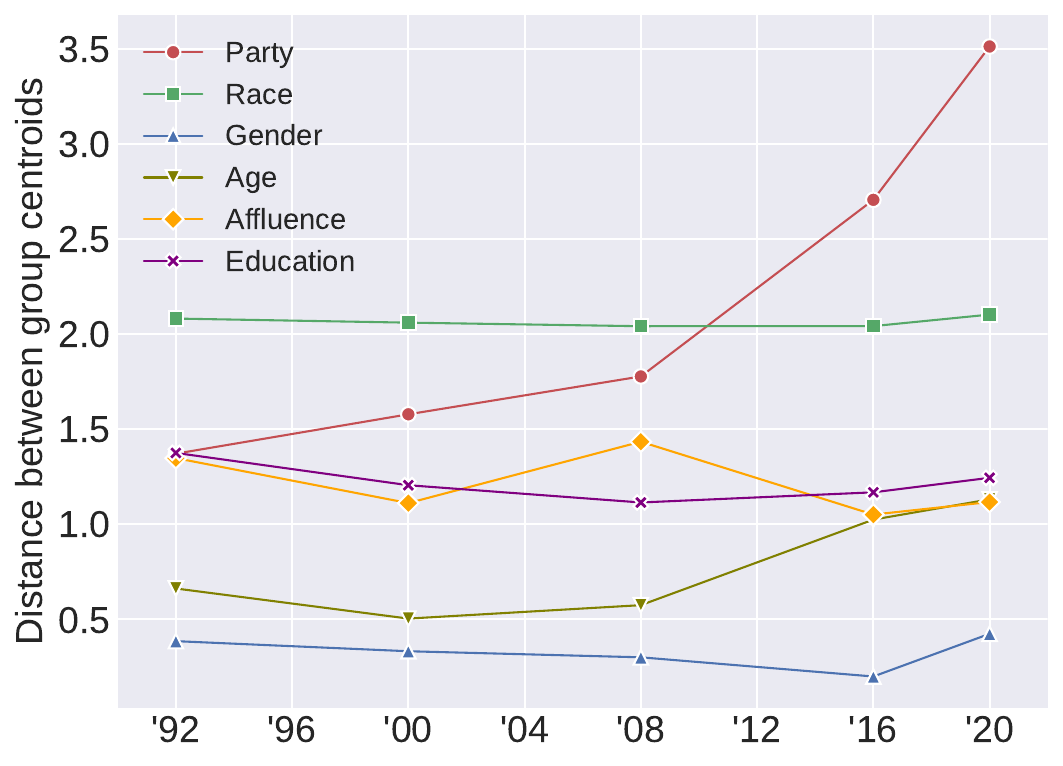}
    \caption{Euclidean distance between the average opinions (centroids) of the opposite groups of each attribute, for every year.
    }
    \label{fig:distances}
\end{figure}

Next, we study the temporal evolution of the opinions of different groups over 30 years. 
To this aim, we start by considering a coarse-grained quantity to capture the aggregate behavior of different socio-economic groups. 
We {compute} the centroid (or geometric center) of each group in each year, defined as the arithmetic mean position of the opinions of the group, indicating the average group opinion.
Fig.~\ref{fig:distances} shows the Euclidean distance between the centroids of opposite groups, for each attribute across the years. 
{We remark that we do not compare distances between different ideological spaces from each year, but, since we select common questions across years, we can chart how different groups' opinions evolve in the same ideological space.
From Fig.~\ref{fig:distances} we see that the groups in Party, Race, Affluence, and Education show larger distances, 
indicating that they are more polarized.
This corroborates the results shown in Figs.~\ref{fig:respondents_density} and \ref{fig:respondents_density_arrows}: 
The attributes with the largest centroid distances correspond to the most ordered opinion distributions with the longest gradient vectors. 
Furthermore, we check that the Euclidean distance between centroids is proportional to the KS distance between distributions, provided by the KS test (see Fig.~7 in the SM). 
This observation confirms that the closer two groups are in the ideological space, the more similar their opinion distributions, a sign of ideological affinity.}

Concerning the temporal evolution, we note that most centroid distances do not vary much over time, with the notable exception of the Party groups: 
The disagreement between Democrats and Republicans has considerably increased over the years, overcoming the ideological separation between racial groups after 2008. 
Notably, this finding is also valid when considering separately policy and non-policy issues (see Fig.~8 in the SM). 
We stress that one cannot directly compare absolute distances between different embeddings (e.g., by comparing the distances reported in Fig.~\ref{fig:distances} and Fig.~8 in the SM), but should only compare distances between different groups in the same embedding. 
As such, we can observe a similar pattern for both policy-related and non-policy-related questions: The groups that show the greatest ideological distance are partisan and racial groups, but only partisan divisions have grown over time. 
The only important difference when splitting between policy and non-policy issues is the larger ideological distance between College and No-College groups for non-policy issues.
We remark that political and demographic groups are not formed by different individuals, but the same individual belongs to many groups, e.g., White, High-Income, College. 
{In order to check that the centroid distances are statistically significant, 
we build a null model assuming that opinions do not differ across groups, see ``Methods'' section ``Null-hypothesis significance testing'' for details. 
We found p-values lower than $10^{-4}$ for all attributes on all years, with the exception of Gender in 2016 ($p = 0.016$). 
This indicates that most socio-economic groups are characterized by different, and in some cases very distant, opinions, as also indicated by the KS test.} 
{Finally, we checked that the ideological distances between opposite groups follow a very similar trend also when considering a three- or four-dimensional ideological space, see Fig.~9 in the SM.}

\begin{figure}[tbp]
    \centering
    \includegraphics[width=0.9\columnwidth]{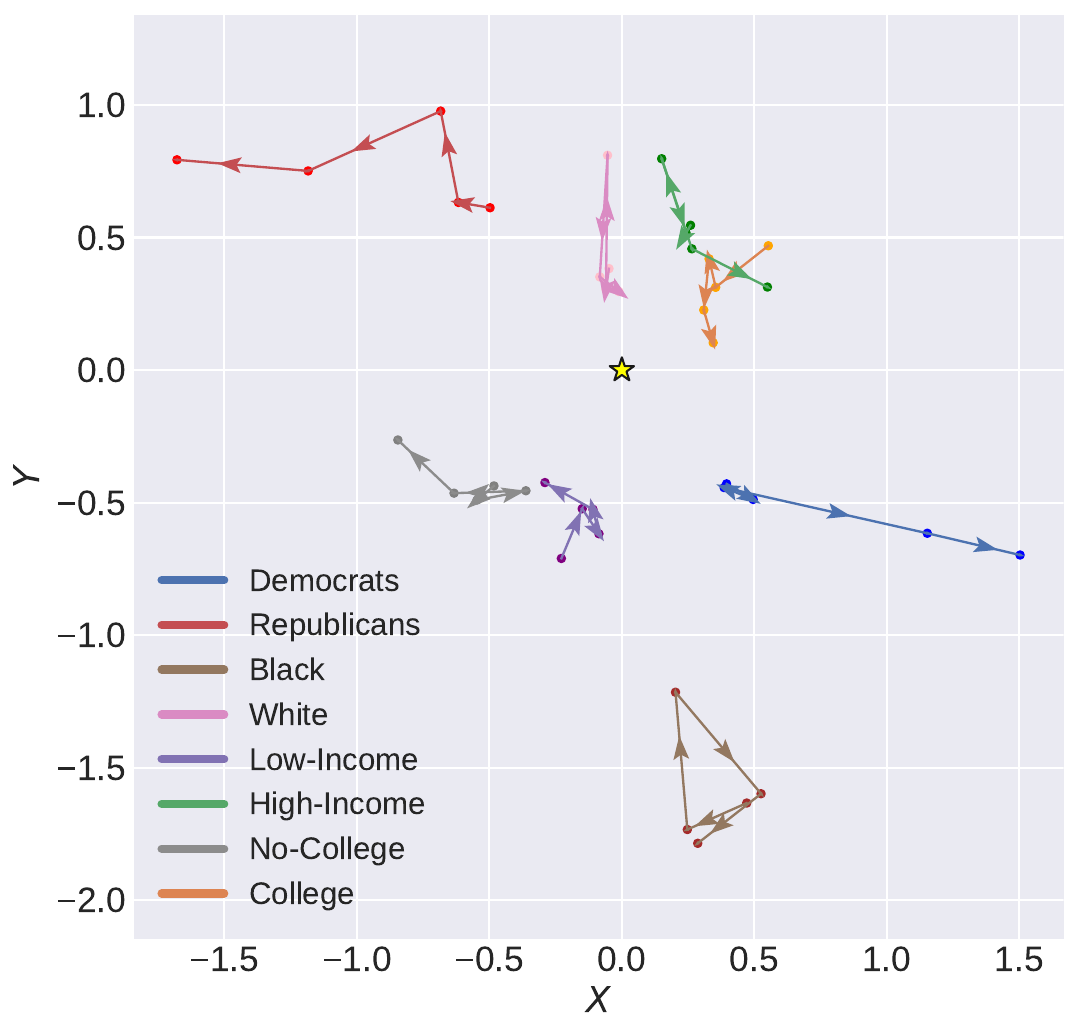}
    \caption{Average opinion (centroid) of the Party, Race, Affluence, and Education groups across years in the ideological space. 
    {The centroids are linked by arrows in chronological sequence: 1992, 2000, 2008, 2016, 2020.} 
    The average opinion of the population of each year is subtracted from each group, being located at $\left( 0, 0 \right)$ and marked with a star. 
    The $X$- and $Y$-axes are arbitrary, combinations of the 29 questions selected.
    }
    \label{fig:centers}
\end{figure}

While the centroid distance provides relevant information regarding the \emph{relative} displacement of the two groups within the ideological space, we are also interested in charting the trajectory of each group over time. 
{With this aim, we study the evolution of the position of the average opinion of the most polarized groups (Party, Race, Affluence, and Education) across years in the ideological space.} 
First, we note that the average opinion of the general population varies over time, especially in 2008 and 2020 (see Fig.~10 in the SM). 
Therefore, to meaningfully compare the trajectory of each group, in Fig.~\ref{fig:centers} we 
discard such a drift by subtracting the average population opinion from each group, for every year. 
In this way, we can see if groups move closer or not to the average population opinion, located at $\left( 0, 0 \right)$ in the Figure.

\begin{figure}[tbp]
    \centering
    \includegraphics[width=0.95\columnwidth]{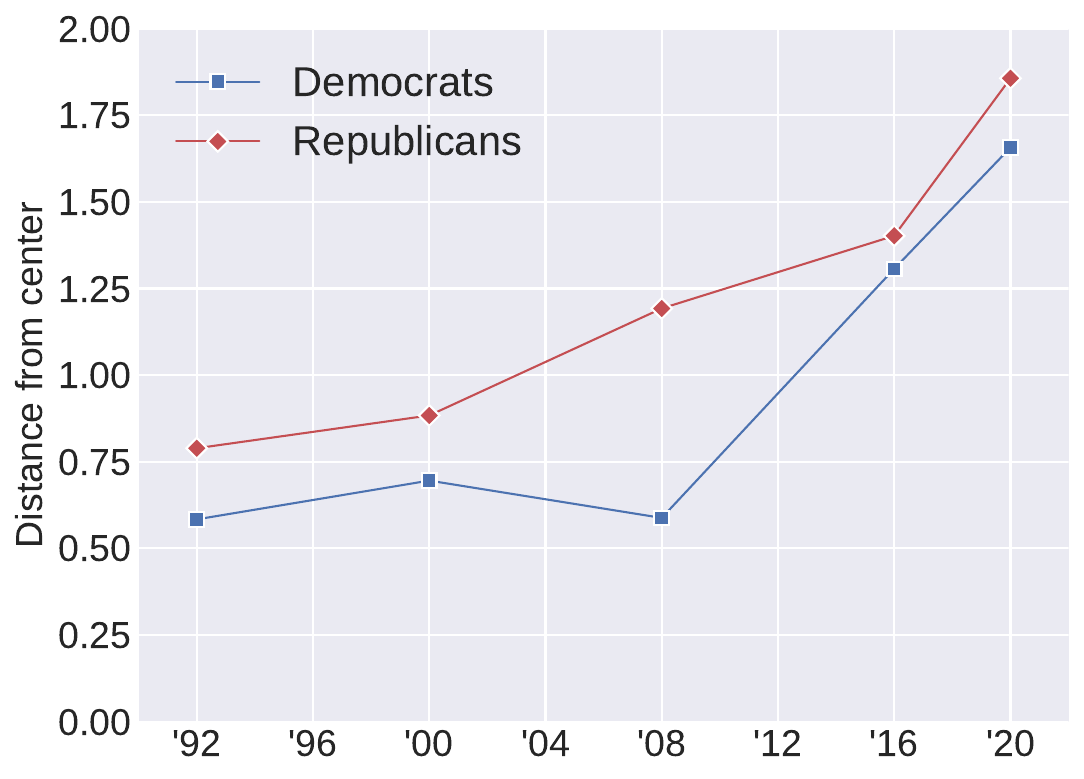}
    \caption{Euclidean distance of the average opinions of Democrats and Republicans from the center, for every year.
    }
    \label{fig:distances_party}
\end{figure}

Note that we recover the spatial disposition of the groups in the ideological space found in Figs.~\ref{fig:respondents_density} and \ref{fig:respondents_density_arrows}: 
Republicans, White, High-Income, and College groups are located in the upper region above the center $\left( 0, 0 \right)$, whereas we find their counterparts at the bottom. 
While most groups orbit in the ideological space without moving much across years, we note two notable exceptions. 
First, the average opinions of the Black group and the general population were particularly similar in 2008. 
This could be the result of the ``Obama Effect''~\cite{welch_obama_2011}, or how Barack Obama's election in 2008 influenced the white prejudice against blacks. 
Between July 2008 and January 2009, racial prejudices were reduced by a rate that was at least five times faster than in the previous two decades~\cite{goldman_effects_2012}. 
The Obama Effect could be reflected in the trajectories charted in the ideological space, with the Black group closer to the general population in 2008. 
Second, the increasing partisan polarization reported in Fig.~\ref{fig:distances} is due to both Republicans and Democrats moving away from the center, especially in 2016 and 2020. 
We address this point more in detail in Fig.~\ref{fig:distances_party}, showing the distance between the Democrats and Republicans from the center each year.
We observe that Republicans are constantly further away from the center than Democrats, as they steadily depart from it since 1992.
In 2016 and 2020, however, Democrats almost caught up with the gap. 
{Fig.~11 in the SM shows that this finding holds when considering a three- or four-dimensional ideological space.}

It is important to bear in mind that the geometric center only provides information about the average opinion of the group. 
One can further characterize the opinion of groups by taking into account the heterogeneity of their distribution. 
If a group populates a large region of the ideological space, their opinions are very heterogeneous, while groups localized within a small region are characterized by more homogeneous opinions. 
In Fig.~12 in the SM we show the radius of gyration of the opinion distribution of all groups considered, for every year {(see ``Methods'' section ``Heterogeneity of the opinion distribution'' for details)}. 
We observe that the opinion dispersion depends more on time than the specific group. 
The radius of gyration, indeed, clearly increased for all groups in 2016 and particularly in 2020. 
In general, opinions within a certain group became more heterogeneous over time, with the exception of Black and Republicans, whose opinions remained more homogeneous.

\begin{figure}[tbp]
    \centering
    \includegraphics[width=0.95\columnwidth]{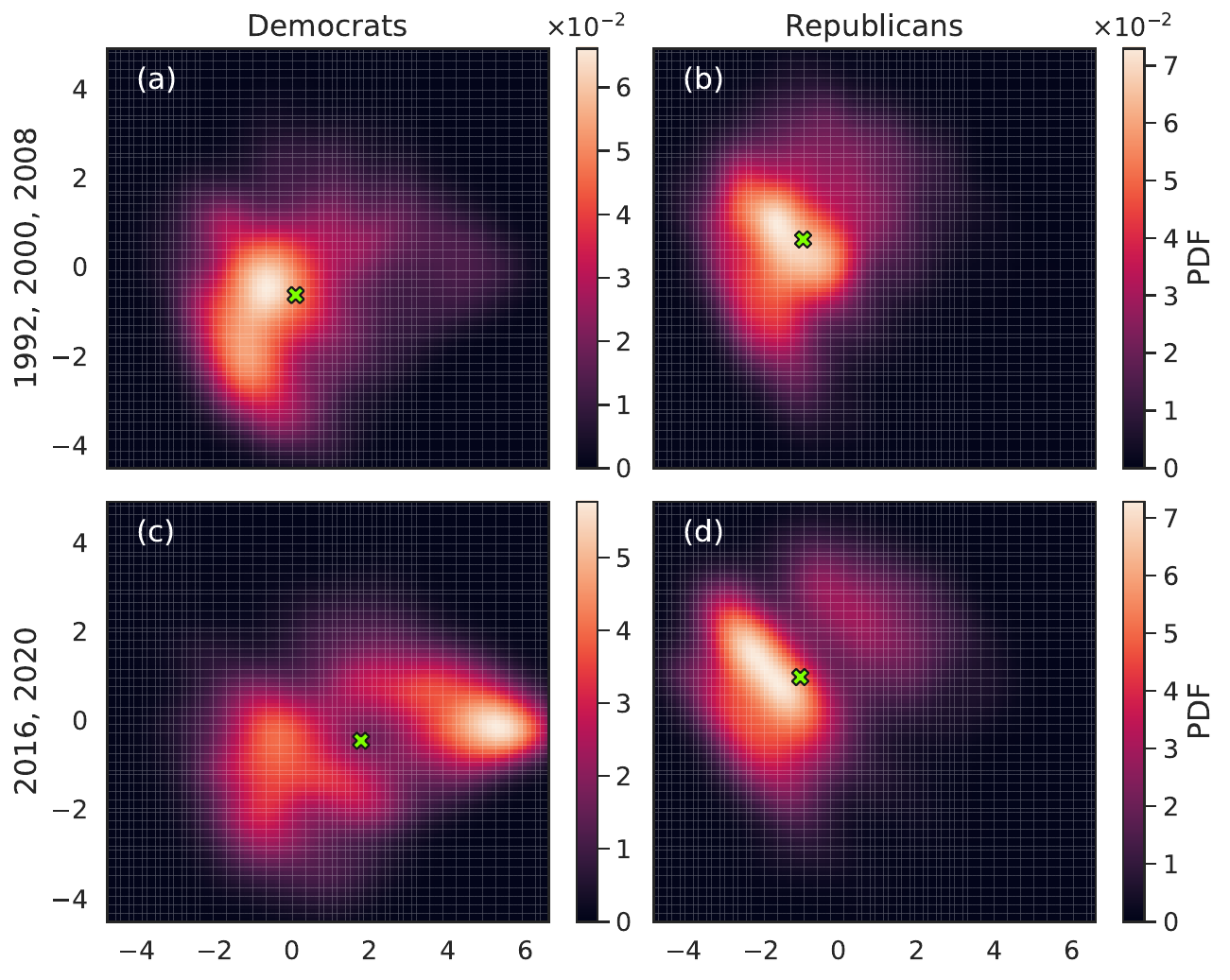}
    \caption{Probability density function (PDF) of Democrats [(a) and (c)] and Republicans [(b) and (d)] in the ideological space. 
    We show the results aggregated over 1992, 2000, and 2008 in (a) and (b), and results over 2016 and 2020 in (c) and (d). 
    {A brighter (darker) color indicates areas with higher (lower) density of individuals.} 
    The green crosses represent the centroid of each distribution. 
    {The axes are arbitrary, combinations of the 29 questions selected.}
    }
    \label{fig:displacement}
\end{figure}

This last observation indicates that 
{not only parties have become more ideologically distant on crucial issues in the last decades, but they also have become less ideologically homogeneous.} 
Fig.~\ref{fig:displacement} supports this intuition, by showing the opinion distribution of Democrats and Republicans aggregated before 2010 (years 1992, 2000, and 2008) and after 2010 (years 2016 and 2020). 
One can see that, while the Republican distribution does not change much from the first 20 years to the last 10, the Democratic distribution has become more dispersed in recent years, extending over a large region of the ideological space. 
{This observation indicates that partisan sorting has declined in the last decade.}

{We quantified the difference observed in Fig.~\ref{fig:displacement} by computing the KS distance between the opinion distributions of different groups. 
We found that the KS distance between Democratic opinions before and after 2010 (panels (a) and (c)) is $0.32$, much higher than the one between Republican opinions in the same time span (panels (b) and (d)), equal to $0.13$. 
For comparison, the KS distance between pre-2010 Democrats and post-2010 Republicans (panels (a) and (d)) is equal to $0.44$.} 
Note that the average opinion (centroid) of Democrats in the years 2016 and 2020, represented by a green cross in Fig.~\ref{fig:displacement}(c), is poorly representative of the whole distribution, as it falls in the middle of the two most populated areas. 
Indeed, a group of respondents emerges in a new region of the space (on the right side of the plots), indicating a clear opinion shift in a share of Democratic voters. 
{By comparing their opinions in the two periods, we discovered that the observed shift is mainly due to stronger opinions regarding racial resentment and other personal attitudes about social justice (see Table~III in the SM). 
To address this point more precisely, we repeat the analysis by splitting the battery into policy-related and non-policy-related questions, finding that the opinion shift among Democrats emerges mainly when considering only questions primarily related to social identity and personal beliefs or attitudes 
(see Fig.~13 in the SM).}

{Furthermore, while racial and income groups are aligned into parties, such an alignment remains roughly constant over time. 
For instance, the fraction of White individuals in the Republican Party was $0.84$ before 2010 and $0.85$ after 2010. In the Democratic Party, such a fraction is $0.56$ and $0.63$, respectively. 
Likewise, the fraction of Low-Income individuals in the Democratic Party was $0.38$ before 2010 and $0.36$ after 2010. 
In the Republican Party, such a fraction is $0.24$ and $0.27$, respectively. 
Therefore, parties became ideologically more distant and less homogeneous, while their alignment with respect to demographic groups remained roughly constant.}

\section*{Discussion}

{The results presented above offer a clear answer to the main research question addressed in the paper, whether ideological polarization actually increased in the last decades, or voters simply sorted themselves into parties matching their ideology more closely. 
By proposing a novel approach, we are able to quantify the evolution of ideological polarization across political and demographic groups over time. 
We found that the ideological distance about fundamental issues between Democrats and Republicans increased in the last 30 years: 
Both parties progressively moved away from the center, at different rates (see Fig.~\ref{fig:distances_party}).
Moreover, we observed that parties, and especially Democrats after 2010, became more heterogeneous: 
{A part of Democrats shifted their opinions regarding several issues, connected to racial resentment and other personal attitudes about social justice (see Fig~\ref{fig:displacement}).} 
These findings directly contradict the hypothesis of partisan sorting as the root of the increasing political polarization in America~\cite{abramowitz_polarized_2013, fiorina2005culture, baldassarri_partisans_2008} ---parties did not become more ideologically distant, they became more homogeneous and coherent on ideological issues.} 
{Furthermore, we explored the partisan alignment of demographic groups, finding that, while some of them (especially racial and income groups) strongly identify with parties (see Fig. \ref{fig:respondents_density_arrows}), such identification does not grow across years. 
Likewise, the ideological distance between all demographic groups did not increase over time, at odds with the ideological distance between parties (see Fig.~\ref{fig:distances}). 
{These findings also offer interesting insights regarding the hypothesis that parties built stronger social identities in the last decades, fueling social polarization~\cite{huddy_expressive_2015, west_partisanship_2022, mason_i_2015}.}

{On the methodological side, the two-dimensional ideological space we defined provides a 
more nuanced characterization of an individual's ideology than the simple one-dimensional liberal/conservative axis, commonly used in the literature.} 
{We remark that our results remain robust across different choices of Isomap hyperparameters, alternative embedding algorithms, and higher dimensionalities of the ideological space.} 
{This modelling choice allows us to chart and easily visualize the evolution of ideological polarization and partisan sorting} 
by computing the distance between opinion distributions and their heterogeneity, as well as the trajectories of political and socio-economic groups in time. 
{Such analysis would not be possible in the traditional one-dimensional representation, or much more cumbersome in higher dimensionalities.} 
Moreover, we selected questions through only objective criteria from the ANES dataset, thus avoiding potential measurement errors resulting from social desirability bias~\cite{brenner_lies_2016}. 
For instance, we excluded all questions related to self-reported scores, like the degree of partisanship or the ideological identity strength. 

It is important to contextualize our findings within the historical period under examination, which witnessed a gradual, and subsequently rapid, realignment of parties along racial, religious, and cultural lines~\cite{abramowitz_ideological_1998, mason_uncivil_2018}. 
It can be argued that the 1992 presidential election marked the inception of the Republican Party's shift toward confrontational politics~\cite{theriault_gingrich_2013}. 
Additionally, 
{the 2008 critical election is often seen as the endpoint of the domination of conservatism in the United States since the late 1970s~\cite{white_barack_2009}.} 
Obama presidency contributed to the swift racialization of American politics~\cite{tesler_spillover_2012}, expanding racial considerations to previously non-racialized issues. 
On the other hand, the 2016 and 2020 elections were characterized by the emergence of a rapidly coalescing anti-diversity coalition of Americans mobilized and organized by Trump~\cite{mason_activating_2021}. 
{The strong partisan divide reflected an unprecedented alignment of racial, religious and ideological issues, which paved the way for the presidency of Trump~\cite{abramowitz_great_2018}.} 
Within this temporal dynamics, it is worth stressing that our methodology captures the ideological transition around the crucial year of 2010, which saw significant shifts in attitudes toward race~\cite{engelhardt_racial_2021}.

Our work is not exempted from some limitations, regarding both the data used and the methodology proposed. Concerning the former, 
the number of respondents in the dataset is not the same every year, e.g., it was particularly low in the year 2000. 
Therefore, we reduced the number of individuals from the more participated surveys, 2016 and 2020, to have a similar number of individuals across years, see ``Methods'' section ``The ANES dataset''. 
Likewise, the number of respondents in opposite groups is similar but not equal: 
Some groups are more populated than their counterparts (see Fig.~3 in the SM). 
The case of the Race attribute is especially noteworthy, since the {majority} of ANES respondents 
{are} White. 
{Furthermore, our methodology requires selecting the same set of questions across time. 
While this choice allows us to embed all individuals in the same ideological space and thus directly compare distances (not possible across different embeddings), it also limits the number and types of questions in the analysis. 
Tracking the same issues over time, indeed, overlooks significant political changes over the past 30 years. 
For instance, it would be important to include questions regarding transgender rights, especially relevant in the last decade.}
{Our dataset includes a mix of policy (anti-gay discrimination law, federal spending, capital punishment), moral values (preference for equality, child-rearing), and procedural (government organization) questions, thus avoiding selecting explicitly partisan-cheerleading questions. 
Instead, we used question-agnostic quantitative criteria that can be applied to other political survey data sets, intending to remove the human bias in hand-picking questions.}

Regarding the methodology instead, a potential issue arises in quantifying ideological polarization based on the distance between average opinions of opposite groups, depicted by centroids in the ideological space. 
This definition may present challenges, particularly when centroids lose significance in highly heterogeneous distributions, such as the one reported in Fig.~\ref{fig:displacement}(c). 
Respondents belonging to the same group could hold substantially different opinions, while the center is located where the density of individuals is almost zero. 
Similarly, the centroid might be influenced by a few extreme opinions, despite the majority of individuals remaining moderate. 
To overcome this limitation, we implemented measures going beyond simple means~\cite{levendusky_red_2011, lelkes_mass_2016}, such as the KS distance measuring overlap between distributions, and the radius of gyration gauging heterogeneity. 
{Furthermore, we obtained fully equivalent results by computing the distribution centroids by means of the median (see Fig.~14 in the SM), which usually works better than the arithmetic mean in the presence of outliers.} 
Finally, we remark that our findings are valid within the boundaries of the ANES dataset used, and limited to the questions selected to build the ideological space.

Despite this latter limitation, we believe the novel methodology proposed here can help political and social scientists to better quantify ideological polarization and social sorting. 
Our framework, indeed, can be applied to any dataset including multiple topics and demographic features of individuals, encompassing a single temporal point or several ones. 
In this latter case, our approach proved particularly useful, allowing us to chart the trajectories of different demographic and political groups into a single ideological space. 
We remark that our methodology does not choose any a priori social or political dimension for the embedding, rather we observed the emergence of two orthogonal axes in the ideological space, corresponding to partisan orientation and education. 
In the future, we hope the ANES will continue to collect data regarding the same (or very similar) political, social, and economic topics, to allow tracking the temporal evolution of demographic groups. 
Assembling data sets rich in the temporal dimension 
is indeed pivotal 
for a longitudinal assessment of political and ideological polarization among the electorate~\cite{abramowitz_disappearing_2010, campbell_polarized_2016, rosenfeld_polarizers_2017}. 
Likewise, with similar data sets it would be possible to gauge more precisely the evolution of social sorting in time. 
{One example is the General Social Survey (GSS)~\cite{GSS}, which has been collecting for 50 years the opinions of Americans on a wide range of topics like confidence in government, abortion or gun rights, together with the demographic information of respondents.} 
Furthermore, our methods could be applied to countries different from the U.S., where 
{both political and societal polarization are threatening governance and democratic norms~\cite{mccoy_polarization_2018}.} 
Finally, while our results are purely observational, future work is needed to assess causal connections between ideological polarization and social sorting.

\section*{Methods}
Here we describe the ANES dataset, the robustness of the dimensionality reduction, the null model testing the significance of the centroids, and the way we compute the heterogeneity of opinion distributions. 
Further details can be found in the Supplementary Material (SM).

\subsection*{The ANES dataset}

The American National Election Studies (ANES)~\cite{anes} is a continuation of a series of academically-run surveys, asking questions to a representative sample of citizens in the United States about their opinions on all sorts of topics. 
The 2022 release includes the answers of 68225 respondents throughout a sample of 32 years from 1948 to 2020. 
The dataset encompasses 1030 variables in total: 161 variables are related to survey information (year, language used, interview mode, etc.) and to information about both interviewer and respondent (gender, race, age, etc.), while the remaining 869 variables consist in the different questions collected by the surveys.

Our objective is to study the ideology of the American population about general topics, thus we narrow down the question list as follows. 
First, we excluded questions about political parties or election candidates (including presidential candidates). 
Then, we excluded binary questions (with only two possible answers), which most of the time refer to single events and not general topics, like ``Do you read a daily newspaper?''. 
{Since not all the questions have the same number of options to answer, we normalized all the scales to the range $\left[ 0,1  \right]$, with negative and positive extreme answers close to $0$ and $1$, respectively.}
The ANES dataset is reduced to a total of 99 questions collecting opinions on different topics such as the state of the economy, government spending, religion, or minority rights.

The questions vary over the years, according to the particular socio-economic situation. 
For example, the disease AIDS was not discovered until the beginning of the eighties, so the question ``Should federal spending on AIDS research be increased?'' was {only} included after 1988.
To compare the opinions collected throughout different years, we can only take into account shared questions. 
Since the number of shared questions between pairs of years is maximum after 1990 (see Fig.~1 in the SM), we focus on the last 30 years. 
Within this time window, we selected the years 1992, 2000, 2008, 2016, and 2020. These years correspond to the election years of the last 5 U.S. Presidents, and {are} also the ones with the largest number of collected questions. 
This leaves us with a total of 41 questions. 
After discarding the ones with a rate of missing answers greater than 20\% (see Fig.~2 in the SM), the number of shared questions is reduced to 29, which are reported in Table~II in the SM. 
Moreover, since the number of respondents in the years 2016 and 2020 is much greater than in 1992, 2000, and 2008, we reduced the former {by choosing a random number of respondents}, to have a similar number of respondents across years. 
{We checked that this random selection of respondents does not alter our results.} 
In chronological sequence, each year finally contains 2485, 1807, 2322, 2500, and 2500 respondents. 
This set with 11614 individuals in total forms the final dataset used in the paper.

\subsection*{Dimensionality reduction}

{To perform the dimensionality reduction of the dataset we apply the Isomap algorithm~\cite{tenenbaum_global_2000}.} 
Isomap considers the distribution of the $K$ neighboring data points by attempting to preserve pairwise geodesic (or curvilinear) distances~\cite{vanderMaaten_dimensionality_2009}. 
Isomap estimates the geodesic distance between data points with the shortest path using Dijkstra's algorithm~\cite{dijkstra_note_1959}. 
The only hyperparameter of the algorithm is then the number $K$ of nearest neighbors. 
All our findings are obtained by setting $K = 10$. 
Figs.~15-17 in the SM show that a different choice of the hyperparameter $K$ leads to fully equivalent results. 
Our Isomap implementation uses \textit{Isomap} from the open-source machine learning library \textit{scikit-learn}~\cite{pedregosa_scikit-learn_2011}.

Regarding the classification task, we relied on cross-validation for the evaluation of the classification performance, and we used logistic regression as the classification algorithm. 
As a strategy to split the data into training and testing sets, we used a Stratified K-Fold with 10 splitting iterations. Our computational implementation uses \textit{cross\_val\_score} and \textit{LogisticRegression} from the library \textit{scikit-learn}~\cite{pedregosa_scikit-learn_2011}.

Furthermore, we tested the robustness of our findings by implementing other embedding algorithms. 
Principal Component Analysis (PCA)~\cite{jolliffe_principal_2002} is a linear dimensionality reduction technique, i.e., the new lower dimensions are a linear combination of the original dimensions. In essence, the low-dimensional representation describes as much of the variance in the high-dimensional data as possible. 
{t-distributed stochastic neighbor embedding (t-SNE)~\cite{vanderMaaten_visualizing_2008} and uniform manifold approximation and projection (UMAP)~\cite{mcinnes2020umap}, on the other hand, are non-linear techniques like Isomap. Both methods are closely related, the low-dimensional space is constructed according to the statistical similarity between original points in the high-dimensional space. 
Figs.~15-17 in the SM show that PCA, t-SNE, and UMAP embeddings lead to similar results. 
We use \textit{PCA} and \textit{TSNE} from \textit{scikit-learn}~\cite{pedregosa_scikit-learn_2011}, and the Python library \textit{umap}~\cite{mcinnes2020umap}.}

\subsection*{Null-hypothesis significance testing}

We tested the statistical significance of the average opinions (centroids) of different demographic groups. 
Our null hypothesis is that opinions do not differ across groups. 
{To test it, we run a bootstrap analysis for each attribute, by randomly assigning each respondent to one of the two opposite groups while preserving their proportion. 
We repeat this process $10^5$ times for each attribute, obtaining the centroids of the demographic groups in every iteration.} 
If each of the resulting $10^5$ centroids per group is given by the real random vector $\mathbf{X} = \left( X, Y \right)$, we assume that it follows a bivariate normal distribution
\begin{equation}
f \left( \mathbf{x} \right) = \frac{1}{2 \pi \sqrt{\left| \Sigma \right|}} \exp{\left[ -\frac{1}{2} \left( \mathbf{x} - \mathbf{\mu} \right) \Sigma^{-1} \left( \mathbf{x} - \mathbf{\mu} \right)^T \right]},
\label{eq:bivariate}
\end{equation}
where $\Sigma$ is the covariance matrix, $\left| \Sigma \right| \equiv \mathrm{det}\left( \Sigma \right)$ and $\mathbf{\mu} = \left( \mu_x, \mu_y \right)$ is the mean vector. The most general form of $\Sigma$ is the symmetric and positive definite matrix
\begin{equation}
\Sigma =
\begin{pmatrix}
\sigma_x^2 & \rho \sigma_x \sigma_y \\
\rho \sigma_x \sigma_y & \sigma_y^2
\end{pmatrix},
\label{eq:covariance}
\end{equation}
where $\sigma_x, \sigma_y$ are the standard deviations of $x$ and $y$, respectively, and $\rho$ is the Pearson correlation coefficient between both coordinates.

By definition, the distance of a point $\mathbf{x} = \left( x,y \right)$ to the two-dimensional distribution given by \eqref{eq:bivariate} reads as
\begin{equation}
r \left( \mathbf{x} \right) = \sqrt{\left( \mathbf{x} - \mathbf{\mu} \right) \Sigma^{-1} \left( \mathbf{x} - \mathbf{\mu} \right)^T},
\label{eq:mahalanobis}
\end{equation}
which is known as the Mahalanobis distance. If we keep it constant, then the location of $\mathbf{x}$ draws an ellipse centered in $\mu$ with semi-major and semi-minor axes given by the greatest and smallest eigenvalues of $\Sigma$ as $r \sqrt{\lambda_1}$ and $r \sqrt{\lambda_2}$, respectively~\cite{ojer_stochastic_2022}. In a polar coordinate system $\left( r, \theta \right)$, with $r$ given by the Mahalanobis distance defined in \eqref{eq:mahalanobis}, a parametric representation of $\mathbf{x}$ can be found as a function of both polar coordinates. Computing analytically the eigenvalues $\lambda_1$ and $\lambda_2$, we write
\begin{eqnarray}
x &=& r \sigma_x \cos{\left( \theta \right)} + \mu_x \nonumber \\
y &=& r \sigma_y \left[ \rho \cos{\left( \theta \right)} + \sqrt{1 - \rho^2} \sin{\left( \theta \right)} \right] + \mu_y.
\label{eq:parametric}
\end{eqnarray}
Thus, its Jacobian determinant is $\mathbf{J}_\mathbf{x} = r \sigma_x \sigma_y \sqrt{1 - \rho^2} = r \sqrt{\left| \Sigma \right|}$.

In a null hypothesis statistical test, under the assumption that the null hypothesis is true, the p-value is defined as the probability of obtaining results equally or more extreme than the result actually observed~\cite{fisher_statistical_1955}. 
{In other words, it measures the probability of obtaining a centroid distribution given by \eqref{eq:bivariate} compatible with the observed centroid.} 
Following this definition, we compute the p-value of a demographic group with opinion centroid at $\mathbf{x} = \mathbf{x}^*$ as
\begin{eqnarray}
p \left( R \right) &=& \frac{1}{2 \pi \sqrt{\left| \Sigma \right|}} \int_0^{2\pi} d\theta \int_R^\infty \mathbf{J}_\mathbf{x} \mathrm{e}^{-\frac{r^2}{2}} dr \nonumber \\
&=& \int_R^\infty r \mathrm{e}^{-\frac{r^2}{2}} dr = \int_{R^2/2}^\infty \mathrm{e}^{-r} dr = \mathrm{e}^{-R^2/2},
\label{eq:p-value}
\end{eqnarray}
where $R \equiv r\left( \mathbf{x}^* \right)$. 
{Our null hypothesis stands that opinions do not differ across groups, which would mean that each observed centroid results from a random distribution, i.e., it is not statistically significant.} 
As usual, we reject the null hypothesis if the p-value is less than or equal to the predefined threshold value 0.05~\cite{moore_introduction_1989}.

\subsection*{Heterogeneity of the opinion distribution}

Each respondent opinion embedded in the low-dimensional space can be written as a data point $\mathbf{x} = \left( x,y \right)$. How heterogeneous is the distribution of opinions can be computed by means of the covariance matrix given by \eqref{eq:covariance}. 
This square matrix is also known as the radius of gyration tensor, since its greatest and smallest eigenvalues $\lambda_1$ and $\lambda_2$, respectively, quantify the mean extension of the distribution in space~\cite{rudnick_shapes_1987, ojer_stochastic_2022}. Commonly, therefore, the radius of gyration is used to describe the average dispersion or heterogeneity, and it is defined as $R_g = \sqrt{\lambda_1 + \lambda_2} = \sqrt{\mathrm{tr} \left( \Sigma \right)} = \sqrt{\sigma_x^2 + \sigma_y^2}$.

\vspace{1cm}

\subsection*{Data and code availability}

The data sets used in this paper are available online\footnote{\url{https://electionstudies.org/data-center/}}.
The code used to analyze data is available online\footnote{\url{https://github.com/polarizationUS/charting_multidimensional_demographic}}.




\begin{acknowledgments}
We thank Corrado Monti and Gianmarco De Francisci Morales for useful comments and discussions. 
J.O and R.P.-S. acknowledge financial support from project PID2022-137505NB-C21/AEI/10.13039/501100011033/ FEDER UE. 
M.S. acknowledges support from Grants No. RYC2022-037932-I and CNS2023-144156 funded by MCIN/AEI/10.13039/501100011033 and the European Union NextGenerationEU/PRTR.
\end{acknowledgments}


\bibliography{NHB}

\clearpage

\newcommand{\thisisthetitle}{Charting multidimensional ideological polarization across demographic groups in the United States}

\pagestyle{fancy}

\lhead{\thisisthetitle\ - SM}
\chead{}
\rhead{}

\lfoot{J. Ojer, D. C\'arcamo, R. Pastor-Satorras, and M. Starnini}
\cfoot{}
\rfoot{\thepage}

\renewcommand{\headrulewidth}{0.5pt}
\renewcommand{\footrulewidth}{0.5pt}

\renewcommand{\thefigure}{\arabic{figure}}
\setcounter{figure}{0}
\renewcommand{\figurename}{Supplementary Figure}
\renewcommand{\thesubfigure}{\roman{subfigure}}
\renewcommand{\tablename}{Supplementary Table}

\title{\thisisthetitle\ \\ \ \\
  Supplementary Material}

\author{Jaume Ojer}
\affiliation{Departament de F\'isica, Universitat Polit\`ecnica de Catalunya, Campus Nord, 08034 Barcelona, Spain}

\author{David C\'arcamo}
\affiliation{Departament de F\'isica, Universitat Polit\`ecnica de Catalunya, Campus Nord, 08034 Barcelona, Spain}

\author{Romualdo Pastor-Satorras}
\affiliation{Departament de F\'isica, Universitat Polit\`ecnica de Catalunya, Campus Nord, 08034 Barcelona, Spain}

\author{Michele Starnini}
\affiliation{Department of Engineering, Universitat Pompeu Fabra, 08018 Barcelona, Spain}
\affiliation{CENTAI Institute, 10138 Turin, Italy}

\maketitle
\onecolumngrid
\begin{center}
\textbf{\large Supplementary Material}
\end{center}

\section{Preprocessing ANES data}

\begin{figure}[h!]
    \centering
    \includegraphics[width=0.65\textwidth]{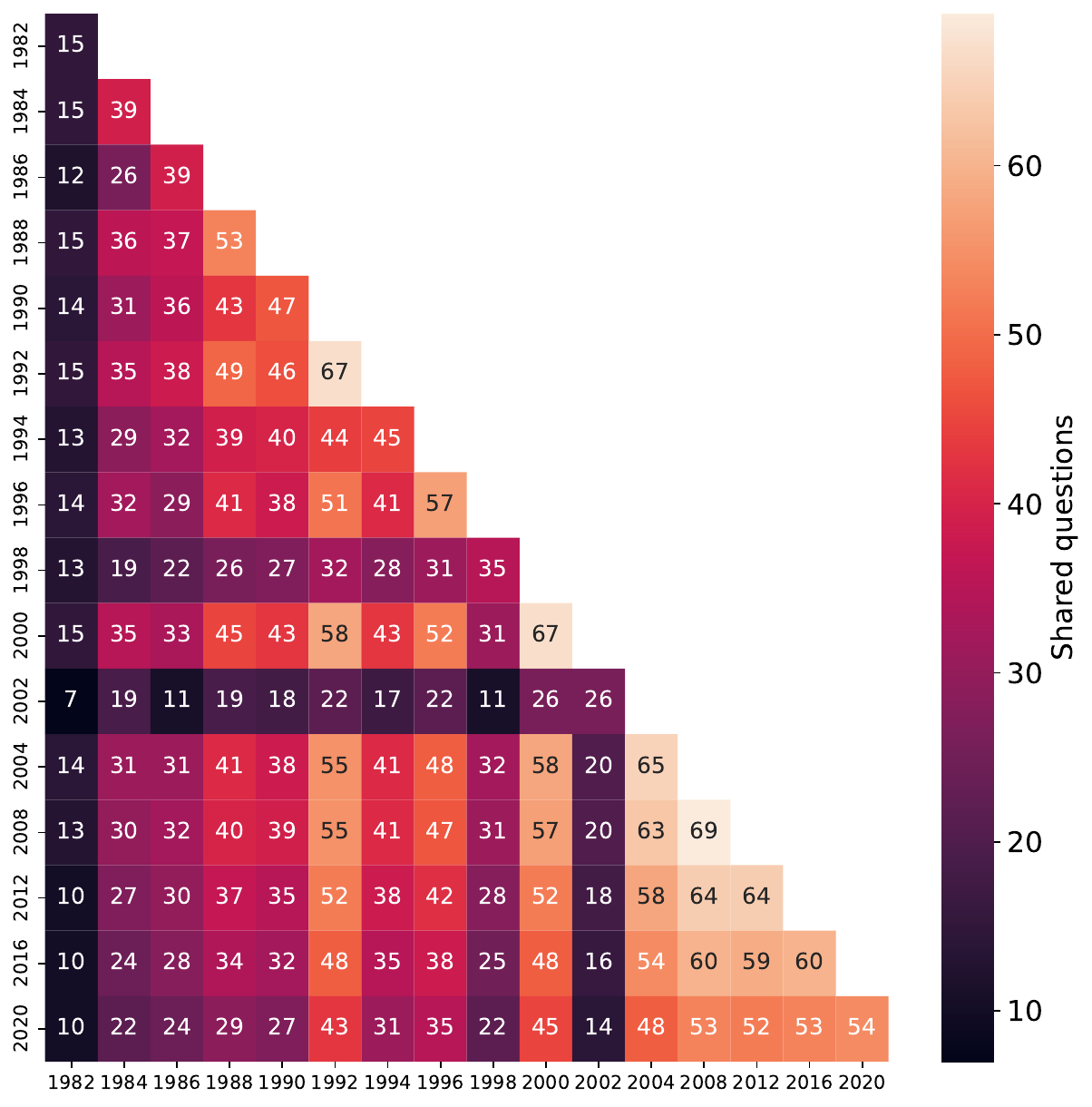}
    \caption{Number of shared questions between pairs of years for ANES editions after 1980. The main diagonal shows the total number of questions collected in every single year regarding the maximum possible of 99 filtered questions.}
    \label{fig:shared_questions}
\end{figure}

In the current release (2022-09-16) of the ANES Cumulative Data File, surveys are performed every 2 to 4 years and conducted in two sessions: before and after presidential elections. 
The number of questions and the questions themselves vary across years depending on the socio-economic situation of the country. 
Supplementary Figure~\ref{fig:shared_questions} shows the number of shared questions between pairs of years from 1982 to 2020. 
In this work, we focus on the ANES editions with the maximum number of questions, which correspond to the last 5 United States presidential elections: 1992, 2000, 2008, 2016, and 2020. 
In Supplementary Table~\ref{tab:shared_questions} we show the number of shared questions for all combinations of the years under consideration. We see that if the number of ANES editions increases or they are more distant in time, the number of available questions decreases.

\begin{table}[tbp]
    \begin{ruledtabular}
        \begin{tabular}{ccc|cc}
        & Years & & Shared questions & \\
        \hline
        & (1992, 2000) & & 58 & \\
        & (1992, 2008) & & 55 & \\
        & (1992, 2016) & & 48 & \\
        & (1992, 2020) & & 43 & \\
        & (2000, 2008) & & 57 & \\
        & (2000, 2016) & & 48 & \\
        & (2000, 2020) & & 45 & \\
        & (2008, 2016) & & 60 & \\
        & (2008, 2020) & & 53 & \\
        & (2016, 2020) & & 53 & \\
        & (1992, 2000, 2008) & & 52 & \\
        & (1992, 2000, 2016) & & 45 & \\
        & (1992, 2000, 2020) & & 41 & \\
        & (1992, 2008, 2016) & & 48 & \\
        & (1992, 2008, 2020) & & 43 & \\
        & (1992, 2016, 2020) & & 43 & \\
        & (2000, 2008, 2016) & & 48 & \\
        & (2000, 2008, 2020) & & 44 & \\
        & (2000, 2016, 2020) & & 44 & \\
        & (2008, 2016, 2020) & & 53 & \\
        & (1992, 2000, 2008, 2016) & & 45 & \\
        & (1992, 2000, 2008, 2020) & & 41 & \\
        & (1992, 2000, 2016, 2020) & & 41 & \\
        & (1992, 2008, 2016, 2020) & & 43 & \\
        & (2000, 2008, 2016, 2020) & & 44 & \\
        & (1992, 2000, 2008, 2016, 2020) & & 41 & \\
        \end{tabular}
    \end{ruledtabular}
    \caption{Number of shared questions available for all of the combinations that result from the ANES editions 1992, 2000, 2008, 2016, 2020.}
    \label{tab:shared_questions}
\end{table}



\subsection{Missing data imputation}

Missing values in the ANES dataset refer to all the answers to specific questions that are not filled in. 
This can be due either because no particular opinion is held, or simply because the respondent does not want to answer. 
Supplementary Figure~\ref{fig:missing} shows that for some of the shared questions, the missing rate among the respondents in each ANES edition is high. 
To work with the most representative samples, we discard all questions with a missing rate greater than 20\%. The total number of shared questions is then reduced to 29.

Roughly, the underlying mechanism to which the data is missing categorizes it into different classes: missing completely at random (MCAR) or missing not at random (MNAR)~\cite{little_statistical_2002}. 
In the context of opinion surveys, the former refers to the case when the probability of having a missing answer does not depend on either the filled answers or the missing answer itself, whereas the latter refers to the case when the data missingness depends equally on both missing and non-missing data. 
However, it is mostly impossible to unambiguously categorize missing data into these mechanisms, since finding if missing data is related or not to other non-missing variables can be very challenging~\cite{graham_missing_2009}. 
Therefore, the most common alternative is to consider the missing data as missing at random (MAR), i.e., independent of the missing values but related to the filled answers~\cite{schafer_analysis_1997}.

There are different treatment methods intended for handling missing data. 
One of the simplest approaches is case deletion, which consists of removing all entries with missing values and considering only the remainers. 
However, this method not only substantially decreases the size of the available dataset, but also introduces bias in data distribution and statistical analysis, especially when missing data is neither MCAR nor MAR. 
Thus, the most popular process is imputation, which involves replacing missing values with an estimate of the distribution of the non-missing values~\cite{little_statistical_2002, donders_review_2006}. 
Simple imputation, for example, replaces missing data by the mean or the median of the available values. 
However, using constants to replace missing data changes the characteristics of the original dataset, which may produce bias and unrealistic results mostly when dealing with high-dimensional datasets. 
Another widely used method is regression imputation, which replaces missing data with predicted values created on a regression model. 
This then assumes that each value depends in some linear way on the other values, although in most cases the relationship is not linear. 
To achieve the highest possible predictive precision, in this work we impute the missing data through machine learning~\cite{emmanuel_survey_2021}. 
In particular, we use the $k$ nearest neighbor ($k$NN) algorithm, by selecting the nearest neighbors of missing values and treating the average value of them as the prediction for imputation. 
For selecting the $k$ nearest neighbors, the similarity between them and missing data should be maximal, i.e., the Euclidean distance is minimized. 
Our implementation uses \textit{KNNImputer} from the open-source machine learning library \textit{scikit-learn}~\cite{pedregosa_scikit-learn_2011}. 
The value selection of the hyperparameter $k$ is probably the most sensitive issue when applying the $k$NN algorithm. 
The substitution of missing data with values that are as close as possible to the true value ceases to be plausible if the data distribution is distorted. 
Although $k > 1$ would in principle lead to better results in terms of imputation precision, 1NN is the only method capable of preserving original data structure~\cite{beretta_nearest_2016}. 
For this reason, in this work, we use 1NN to impute all the missing answers.

\begin{figure}[tbp]
    \centering
    \includegraphics[width=0.9\textwidth]{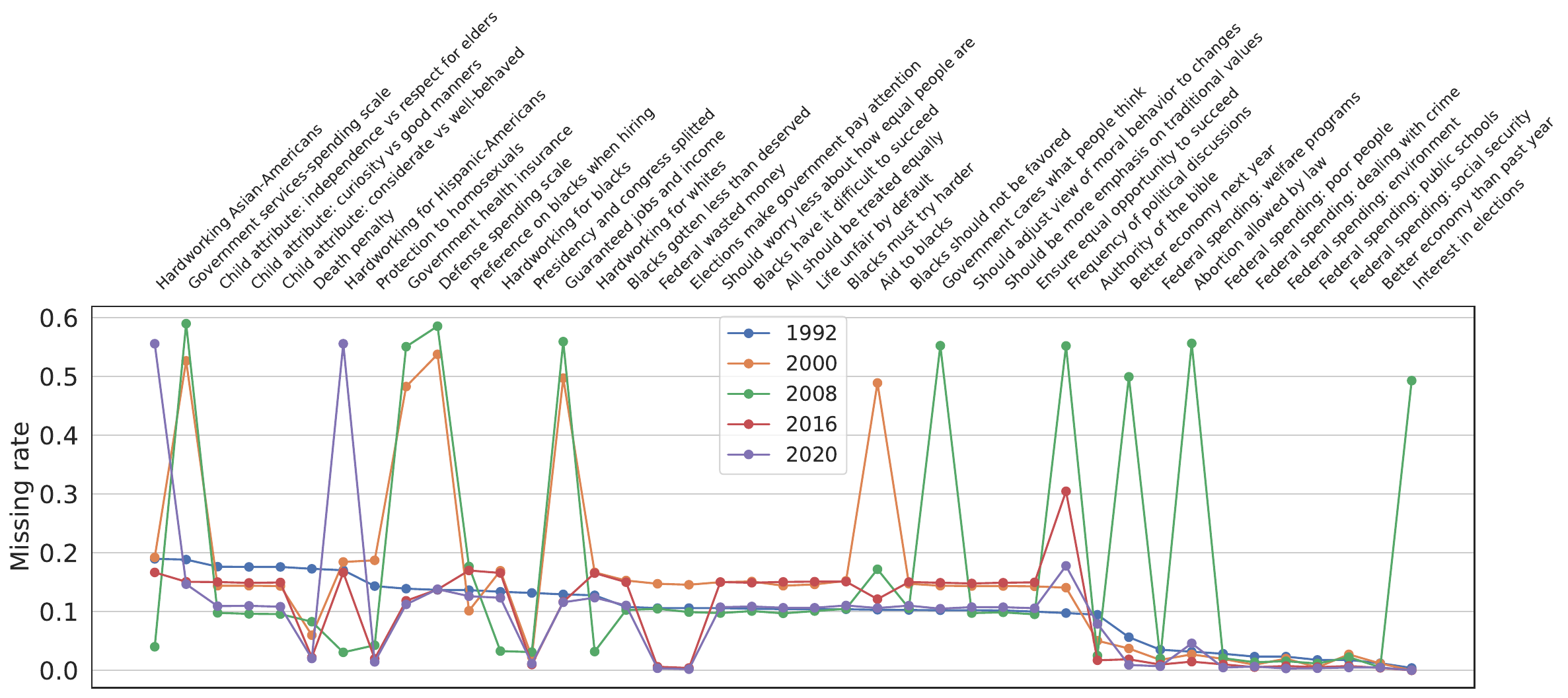}
    \caption{Rate of missing data for each of the 41 shared questions within the years 1992, 2000, 2008, 2016, and 2020. For our work, we considered the 29 questions with a missing rate lower than 20\%.
    }
    \label{fig:missing}
\end{figure}


\begin{table} [b!]
    \centering
    \begin{tabular}{|p{5cm}|p{10cm}|p{2cm}|}
        \hline
        \textbf{Issue} & \textbf{Question} & \textbf{ANES ID} \\
        \hline
        Federal wasted money & Do you think that people in the government waste a lot of money we pay in taxes, waste some of it, or don’t waste very much of it? \newline [A lot, Some, Not very much] & VCF0606 \\
        \hline
        Elections make government pay attention & How much do you feel that having elections makes the government pay attention to what the people think? \newline [Not much, Some, A good deal] & VCF0624 \\
        \hline
        Authority of the Bible & Which of these statements comes closest to describing your feelings about the Bible? \newline [The Bible is the actual Word of God and is to be taken literally, word for word; The Bible is the Word of God but not everything in it should be taken literally, word for word; The Bible is a book written by men and is not the Word of God] & VCF0850 \\
        \hline
        Should adjust view of moral behavior to changes & ‘The world is always changing and we should adjust our view of moral behavior to those changes.’ \newline [Agree strongly, Agree somewhat, Neither agree nor disagree, Disagree somewhat, Disagree strongly] & VCF0852 \\
        \hline
        Should be more emphasis on traditional values & ‘This country would have many fewer problems if there were more emphasis on traditional family ties.’ \newline [Agree strongly, Agree somewhat, Neither agree nor disagree, Disagree somewhat, Disagree strongly] & VCF0853 \\
        \hline
        Preference on blacks when hiring & Are you for or against preferential hiring and promotion of blacks? \newline [Favor strongly, Favor not strong, Oppose not strong, Oppose strongly] & VCF0867a \\
        \hline
        Better economy than past year & Would you say that over the past year the nation’s economy has gotten better, stayed the same or gotten worse? \newline [Much better, Somewhat better, Stayed same, Somewhat worse, Much worse] & VCF0871 \\
        \hline
        Protection to homosexuals & Do you favor or oppose laws to protect homosexuals/gays and lesbians against job discrimination? \newline [Favor strongly, Favor not strong, Oppose not strong, Oppose strongly] & VCF0876a \\
        \hline
        Federal spending: \newline poor people & Should federal spending on aid to poor people be increased, decreased, or kept about the same? \newline [Increased, Same, Decreased or cut out entirely] & VCF0886 \\
        \hline
        Federal spending: \newline dealing with crime & Should federal spending on dealing with crime be increased, decreased, or kept about the same? \newline [Increased, Same, Decreased or cut out entirely] & VCF0888 \\
        \hline
        Federal spending: \newline public schools & Should federal spending on public schools be increased, decreased, or kept about the same? \newline [Increased, Same, Decreased or cut out entirely] & VCF0890 \\
        \hline
    \end{tabular}
\end{table}

\begin{table} [tbp]
    \centering
    \begin{tabular}{|p{5cm}|p{10cm}|p{2cm}|}
        \hline
        Federal spending: \newline welfare programs & Should federal spending on welfare programs be increased, decreased, or kept about the same? \newline [Increased, Same, Decreased or cut out entirely] & VCF0894 \\
        \hline
        Ensure equal opportunity to succeed & ‘Our society should do whatever is necessary to make sure that everyone has an equal opportunity to succeed.’ \newline [Agree strongly, Agree somewhat, Neither agree nor disagree, Disagree somewhat, Disagree strongly] & VCF9013 \\
        \hline
        Life unfair by default & ‘It is not really that big a problem if some people have more of a chance in life than others.’ \newline [Agree strongly, Agree somewhat, Neither agree nor disagree, Disagree somewhat, Disagree strongly] & VCF9016 \\
        \hline
        Should worry less about how equal people are & ‘This country would be better off if we worried less about how equal people are.’ \newline [Agree strongly, Agree somewhat, Neither agree nor disagree, Disagree somewhat, Disagree strongly] & VCF9017 \\
        \hline
        All should be treated equally & ‘If people were treated more equally in this country we would have many fewer problems.’ \newline [Agree strongly, Agree somewhat, Neither agree nor disagree, Disagree somewhat, Disagree strongly] & VCF9018 \\
        \hline
        Blacks have it difficult to succeed & ‘Generations of slavery and discrimination have created conditions that make it difficult for blacks to work their way out of the lower class.’ \newline [Agree strongly, Agree somewhat, Neither agree nor disagree, Disagree somewhat, Disagree strongly] & VCF9039 \\
        \hline
        Blacks should not be favored & ‘Irish, Italians, Jewish and many other minorities overcame prejudice and worked their way up. Blacks should to the same without any special favors.’ \newline [Agree strongly, Agree somewhat, Neither agree nor disagree, Disagree somewhat, Disagree strongly] & VCF9040 \\
        \hline
        Blacks must try harder & ‘It’s really a matter of some people not trying hard enough; if blacks would only try harder they could be just as well off as whites.’ \newline [Agree strongly, Agree somewhat, Neither agree nor disagree, Disagree somewhat, Disagree strongly] & VCF9041 \\
        \hline
        Blacks gotten less than deserved & ‘Over the past few years blacks have gotten less than they deserve.’ \newline [Agree strongly, Agree somewhat, Neither agree nor disagree, Disagree somewhat, Disagree strongly] & VCF9042 \\
        \hline
        Federal spending: \newline environment & Should federal spending on improving and protecting the environment be increased, decreased, or stay the same? \newline [Increased, Same, Decreased] & VCF9047 \\
        \hline
        Federal spending: \newline social security & Should federal spending on social security be increased, decreased, or stay the same? \newline [Increased, Same, Decreased] & VCF9049 \\
        \hline
        Presidency and congress splitted & Do you think it is better when one party controls both the presidency and Congress, or when control is split between the Democrats and Republicans? \newline [One party control both, Control is split, It doesn’t matter] & VCF9206 \\
        \hline
        Death penalty & Do you favor or oppose the death penalty for persons convicted of murder? \newline [Favor strongly, Favor not strong, Oppose not strong, Oppose strongly] & VCF9237 \\
        \hline
        Child attribute: \newline curiosity vs good manners & Which one you think is more important for a child to have? \newline [Curiosity, Both, Good manners] & VCF9246 \\
        \hline
        Child attribute: \newline considerate vs well-behaved & Which one you think is more important for a child to have? \newline [Being considerate, Both, Well behaved] & VCF9248 \\
        \hline
        Child attribute: \newline independence vs respect for elders & Which one you think is more important for a child to have? \newline [Independence, Both, Respect for elders] & VCF9249 \\
        \hline
        Hardworking for whites & Where would you rate whites on a scale of 1 to 7? (where 1 indicates hard working, 7 means lazy, and 4 indicates most whites are not closer to one end or the other.) \newline [1. Hard-working - 7. Lazy] & VCF9270 \\
        \hline
        Hardworking for blacks & Where would you rate blacks on a scale of 1 to 7? (where 1 indicates hard working, 7 means lazy, and 4 indicates most blacks are not closer to one end or the other.) \newline [1. Hard-working - 7. Lazy] & VCF9271 \\
        \hline
    \end{tabular}
    \caption{Overview of all 29 questions and their respective ANES IDs.}
    \label{tab:ANESquestions}
\end{table}

\clearpage

\section{Supplementary Figures}

\begin{figure}[h!]
    \centering
    \includegraphics[width=0.75\textwidth]{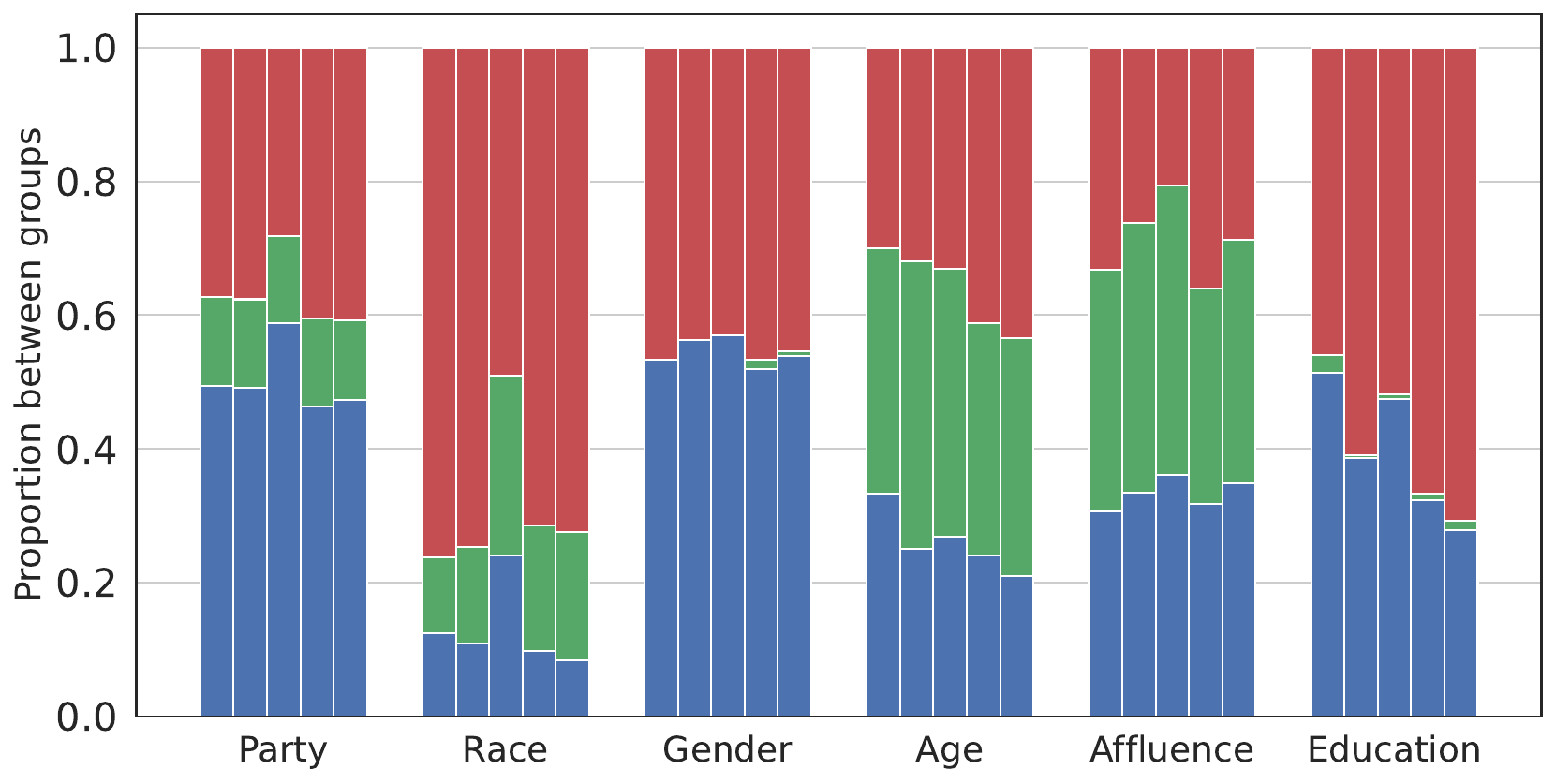}
    \caption{Proportion of individuals belonging to different socio-economic groups of all attributes considered, over the years (ordered from left to right for each attribute). Following Figure~1 of the main text, we colored in blue (red) the groups of Democrats, Black, Female, 17-34, Low-Income, and No-College (Republicans, White, Male, 55+, High-Income, and College). We colored in green the remaining groups that are not considered in our work. The total number of individuals is 2485, 1807, 2322, 2500, and 2500 for the years 1992, 2000, 2008, 2016, and 2020, respectively.
    }
    \label{fig:proportions}
\end{figure}

\begin{figure}[h!]
    \centering
    \includegraphics[width=0.6\columnwidth]{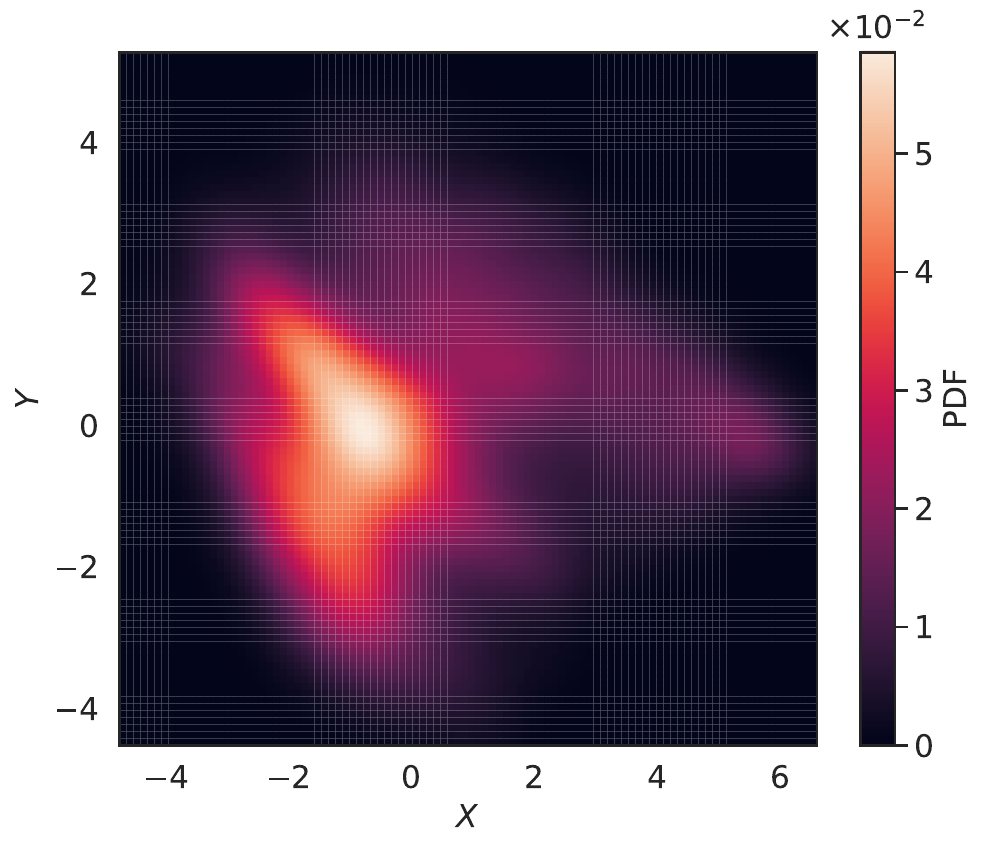}
    \caption{Probability density function (PDF) of the ANES respondents in the two-dimensional ideological space obtained by Isomap with $K = 10$. 
    The lighter (darker) areas are the ones with greater (lower) density. 
    {The $X$- and $Y$-axes of the ideological space are arbitrary, combinations of the 29 questions selected.}
    }
    \label{fig:embedding}
\end{figure}

\clearpage

\begin{figure}[h!]
    \centering
    \includegraphics[width=0.65\columnwidth]{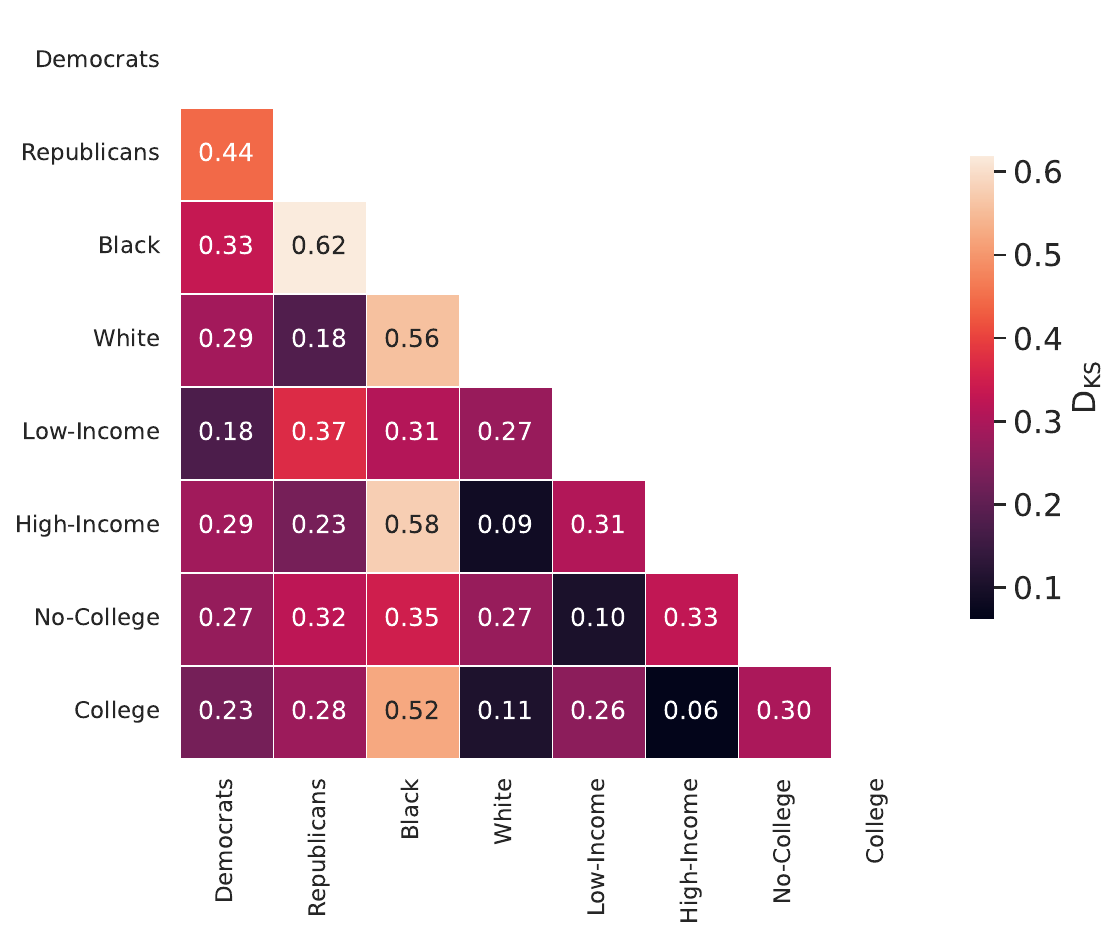}
    \caption{KS distance $D_{KS}$ between different socio-economic groups of the most polarized attributes, aggregated over time. We made the computations by using an updated Python implementation given by the \textit{ndtest} package~\cite{syrte_ndtest}.
    }
    \label{fig:KS_groups}
\end{figure}

\begin{figure}[h!]
    \centering
    \includegraphics[width=0.45\columnwidth]{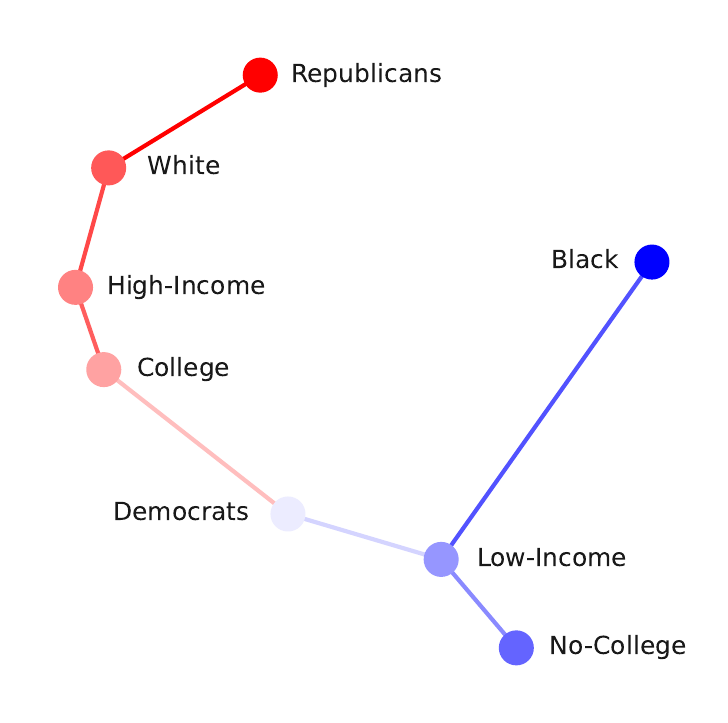}
    \caption{Minimum spanning tree (MST) of the ideological affinity among groups of Party, Race, Affluence, and Education. 
    The MST is constructed by connecting pairs of groups with edges, such that all groups are connected, there are no cycles, and the total KS distance between groups joined by edges is the minimum possible. 
    The length of the edges is proportional to the KS distance between nodes. The two nodes with the largest distance (Black and Republicans) are colored with the darkest colors, while the rest of the nodes are colored according to their distance from these two groups. 
    The position of the nodes is arbitrary.
    }
    \label{fig:MST}
\end{figure}

\begin{figure}[h!]
    \centering
    \includegraphics[width=0.55\columnwidth]{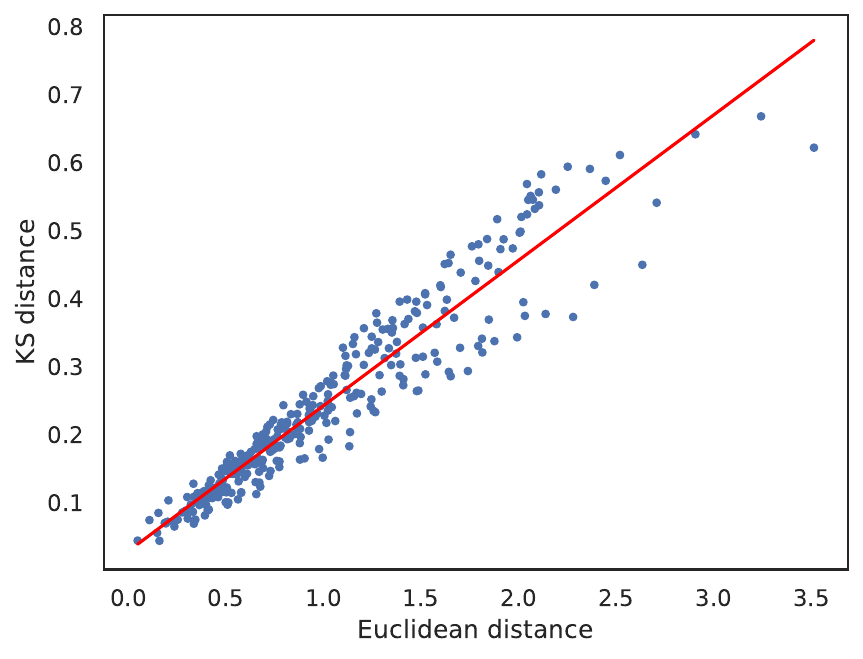}
    \caption{KS distance as a function of the Euclidean distance between the centroids of all socio-economic groups considered. Each point corresponds to one year among 1992, 2000, 2008, 2016, and 2020. The red line represents the linear regression between both distances, leading to a Pearson correlation coefficient of $r = 0.95$.
    }
    \label{fig:KS_Euclidean}
\end{figure}

\begin{figure}[h!]
\subfloat[]{
  \includegraphics[width=0.47\textwidth]{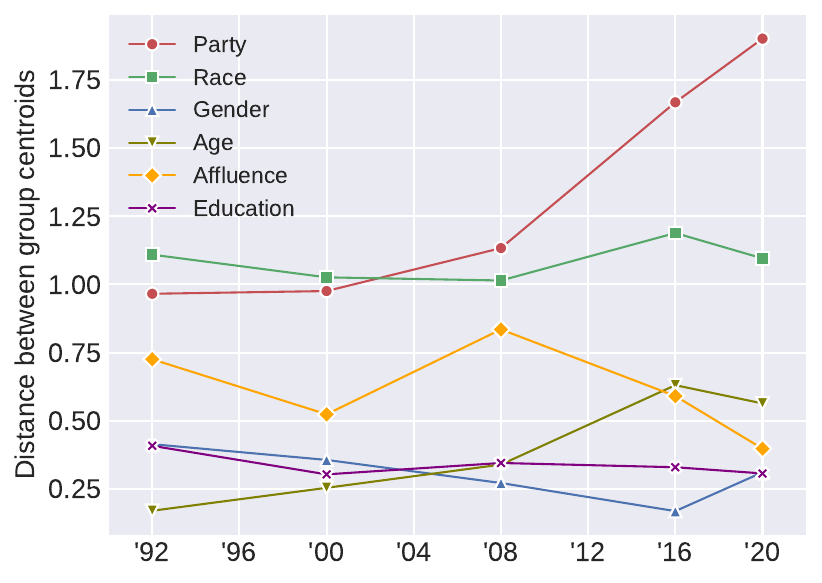}
}
\subfloat[]{
  \includegraphics[width=0.47\textwidth]{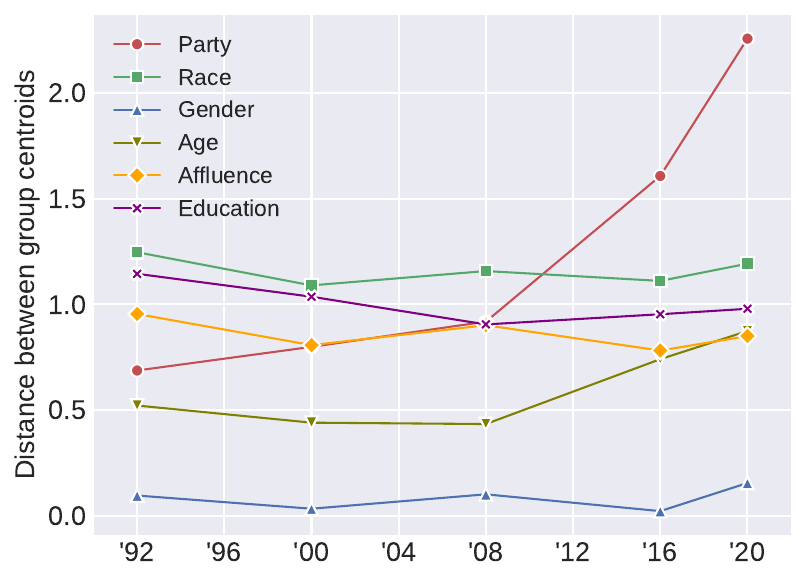}
}
\caption{
Euclidean distance between the distribution centroids of opposite groups for every year, by considering (i) opinions with respect to policy, and (ii) opinions with respect to social identity and personal beliefs/attitudes. 
We consider questions 1, 6, 8-13, 21, 22, 24 (in the order of Table~\ref{tab:ANESquestions}) as policy-related, and questions 2-5, 7, 14-20, 23, 25-29 as non-policy-related.
}
\label{fig:distances_policyVSnonpolicy}
\end{figure}

\begin{figure}[h!]
\subfloat[]{
  \includegraphics[width=0.47\textwidth]{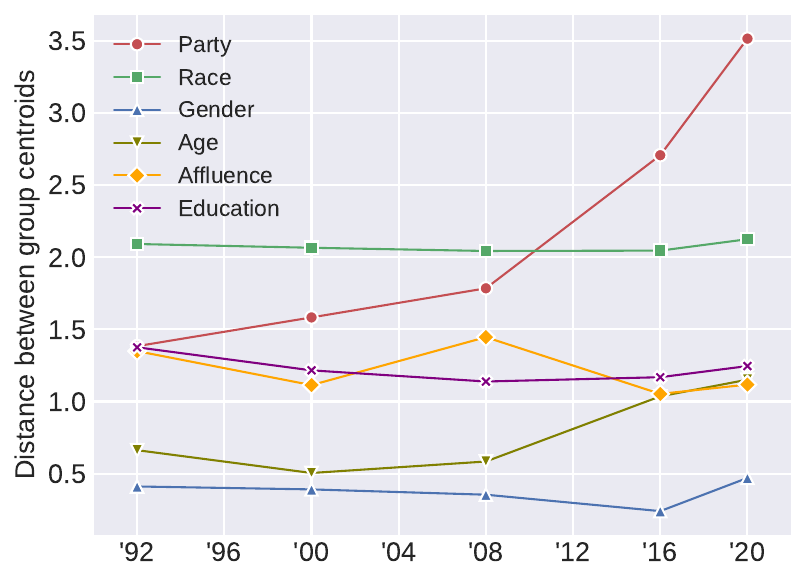}
}
\subfloat[]{
  \includegraphics[width=0.47\textwidth]{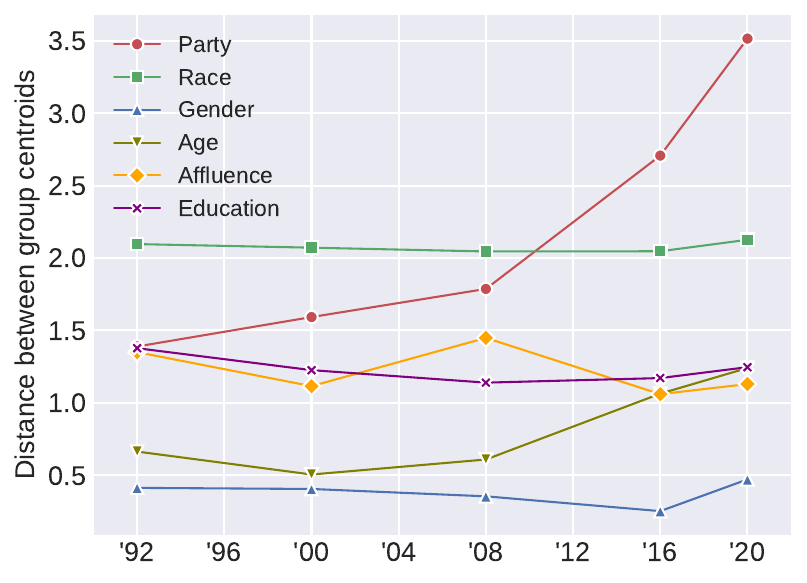}
}
\caption{
Euclidean distance between the distribution centroids of opposite groups for every year, by embedding the ANES opinions in (i) 3 dimensions, and (ii) 4 dimensions.
}
\label{fig:distances_3Dvs4D}
\end{figure}

\begin{figure}[h!]
    \centering
    \includegraphics[width=0.55\columnwidth]{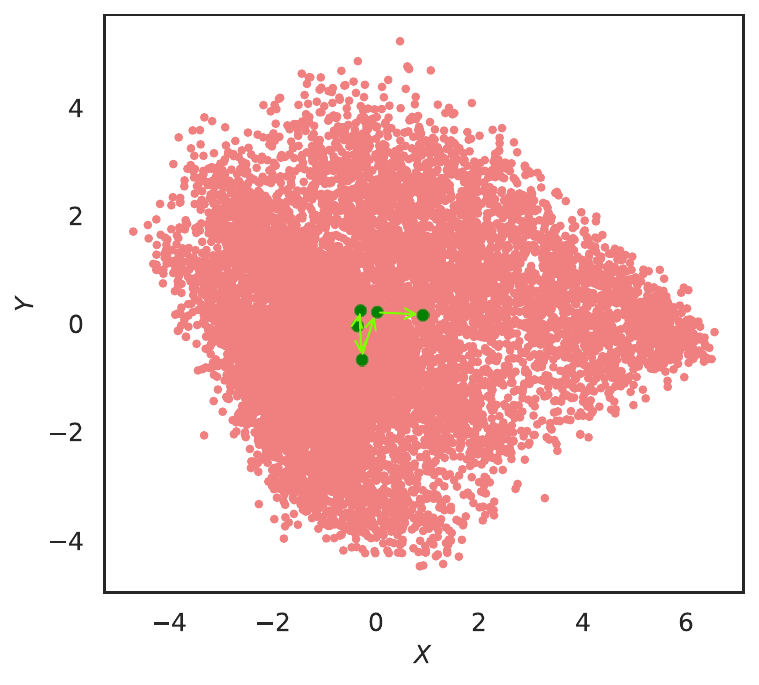}
    \caption{Two-dimensional ideological space, where each point corresponds to the embedded opinion of each ANES respondent. 
    The average opinions of the different years (1992, 2000, 2008, 2016, 2020) are linked by arrows in chronological sequence. 
    The $X$- and $Y$-axes of the ideological space are arbitrary, combinations of the 29 questions selected.
    }
    \label{fig:embedding_arrows}
\end{figure}

\begin{figure}[h!]
\subfloat[]{
  \includegraphics[width=0.47\textwidth]{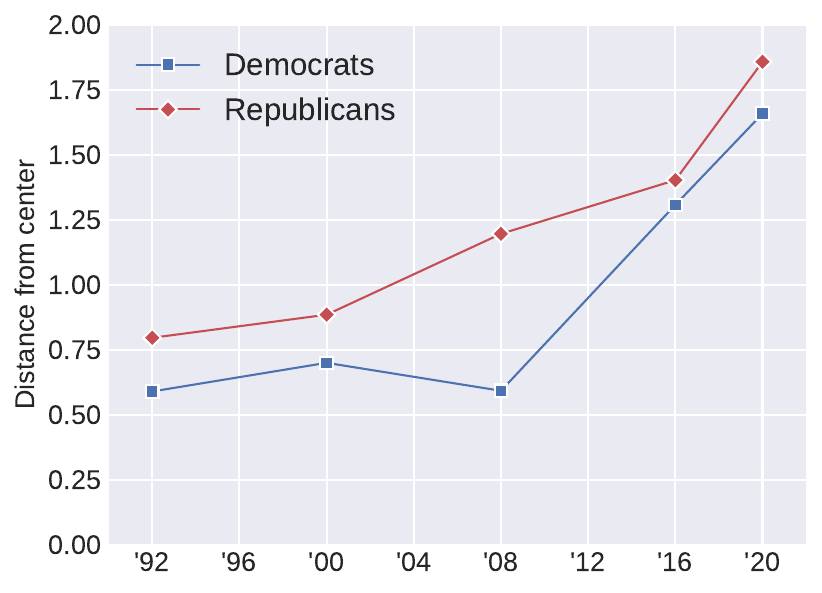}
}
\subfloat[]{
  \includegraphics[width=0.47\textwidth]{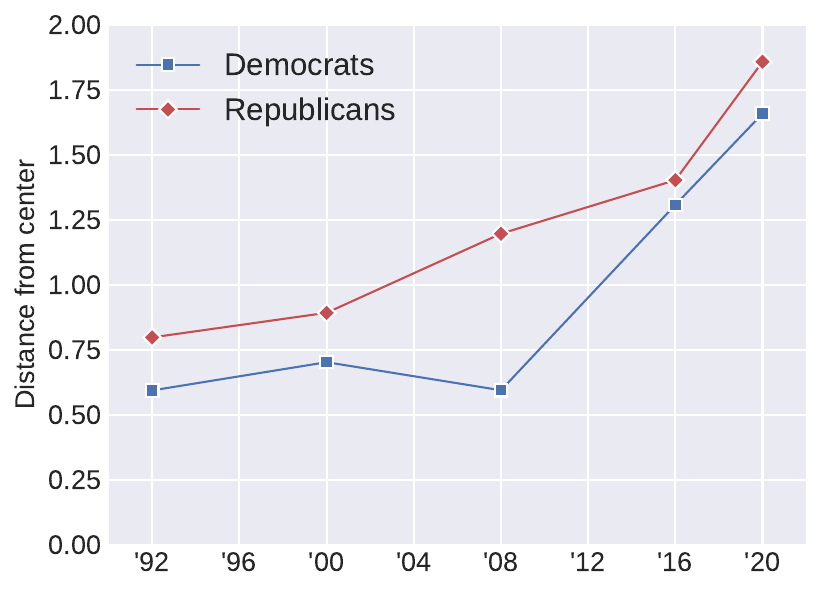}
}
\caption{
Euclidean distance of the average opinions of Democrats and Republicans from the center for every year, by embedding the ANES opinions in (i) 3 dimensions, and (ii) 4 dimensions.
}
\label{fig:distances_party_3Dvs4D}
\end{figure}

\begin{figure}[h!]
    \centering
    \includegraphics[width=0.75\columnwidth]{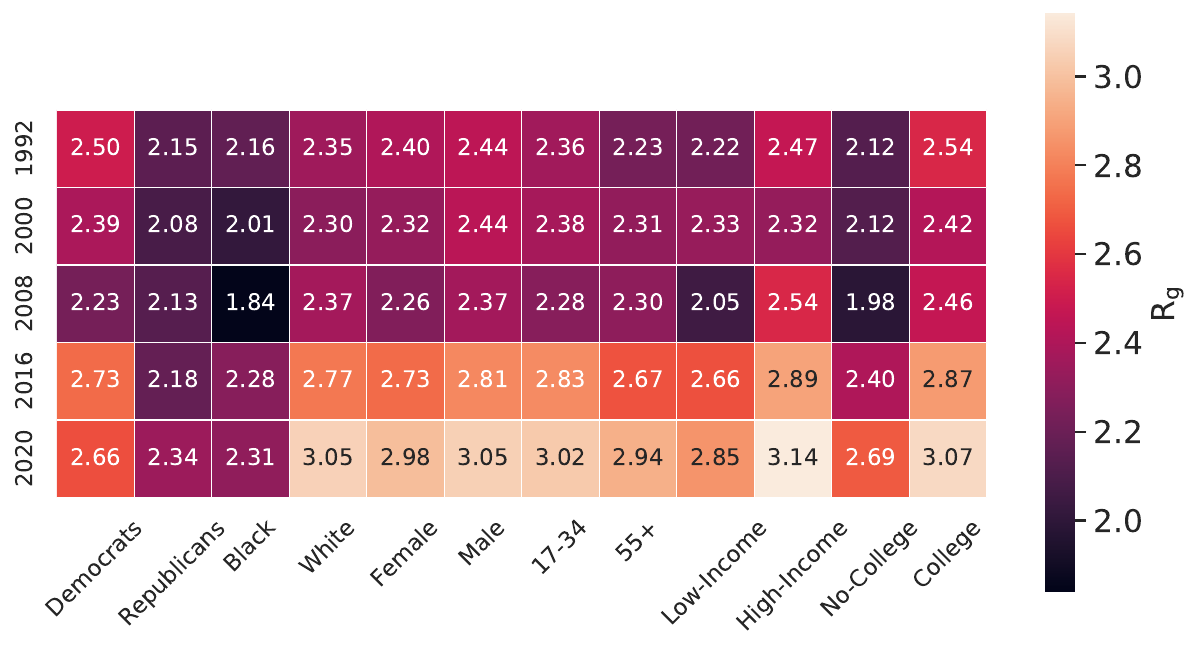}
    \caption{Radius of gyration $R_g$ of the opinion distributions for each group and year.}
    \label{fig:radius_gyration}
\end{figure}

\begin{figure}[h!]
\subfloat[]{
  \includegraphics[width=0.47\textwidth]{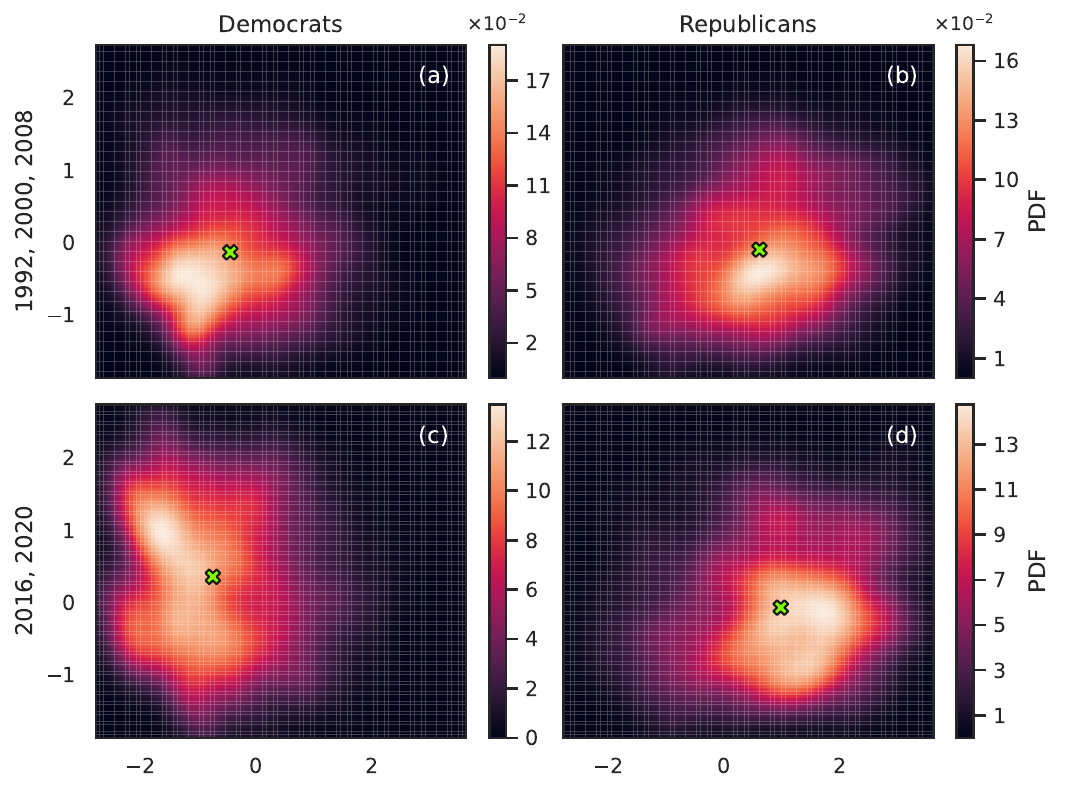}
}
\subfloat[]{
  \includegraphics[width=0.47\textwidth]{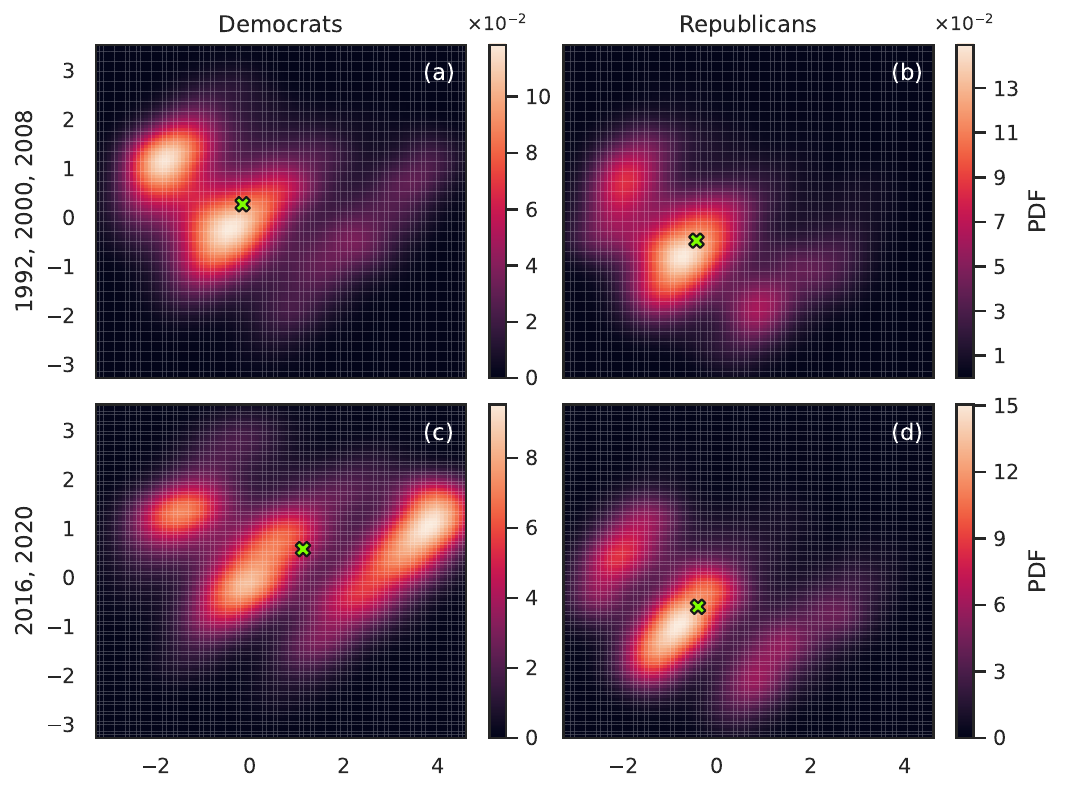}
}
\caption{
Probability density function of Democrats and Republicans aggregated over 1992, 2000, and 2008, and over 2016 and 2020, by considering (i) opinions with respect to policy, and (ii) opinions with respect to social identity and personal beliefs/attitudes. 
The green crosses represent the centroid of each distribution. 
We consider questions 1, 6, 8-13, 21, 22, 24 (in the order of Table~\ref{tab:ANESquestions}) as policy-related, and questions 2-5, 7, 14-20, 23, 25-29 as non-policy-related.
}
\label{fig:displacement_policyVSnonpolicy}
\end{figure}

\begin{figure}[h!]
    \centering
    \includegraphics[width=0.55\columnwidth]{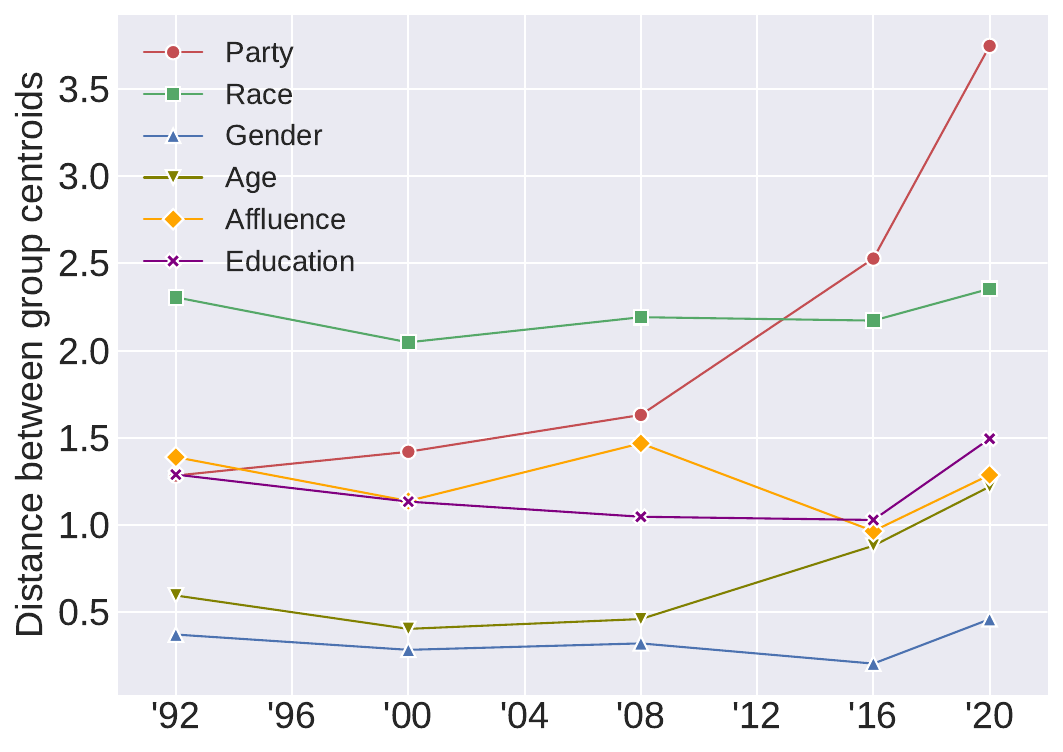}
    \caption{
    Euclidean distance between the distribution centroids of opposite groups computed by the median of opinions, for every year.
    }
    \label{fig:distances_median}
\end{figure}

\clearpage

\section{Robustness of results}

\begin{figure}[h!]
\subfloat[]{
  \includegraphics[width=0.4\textwidth]{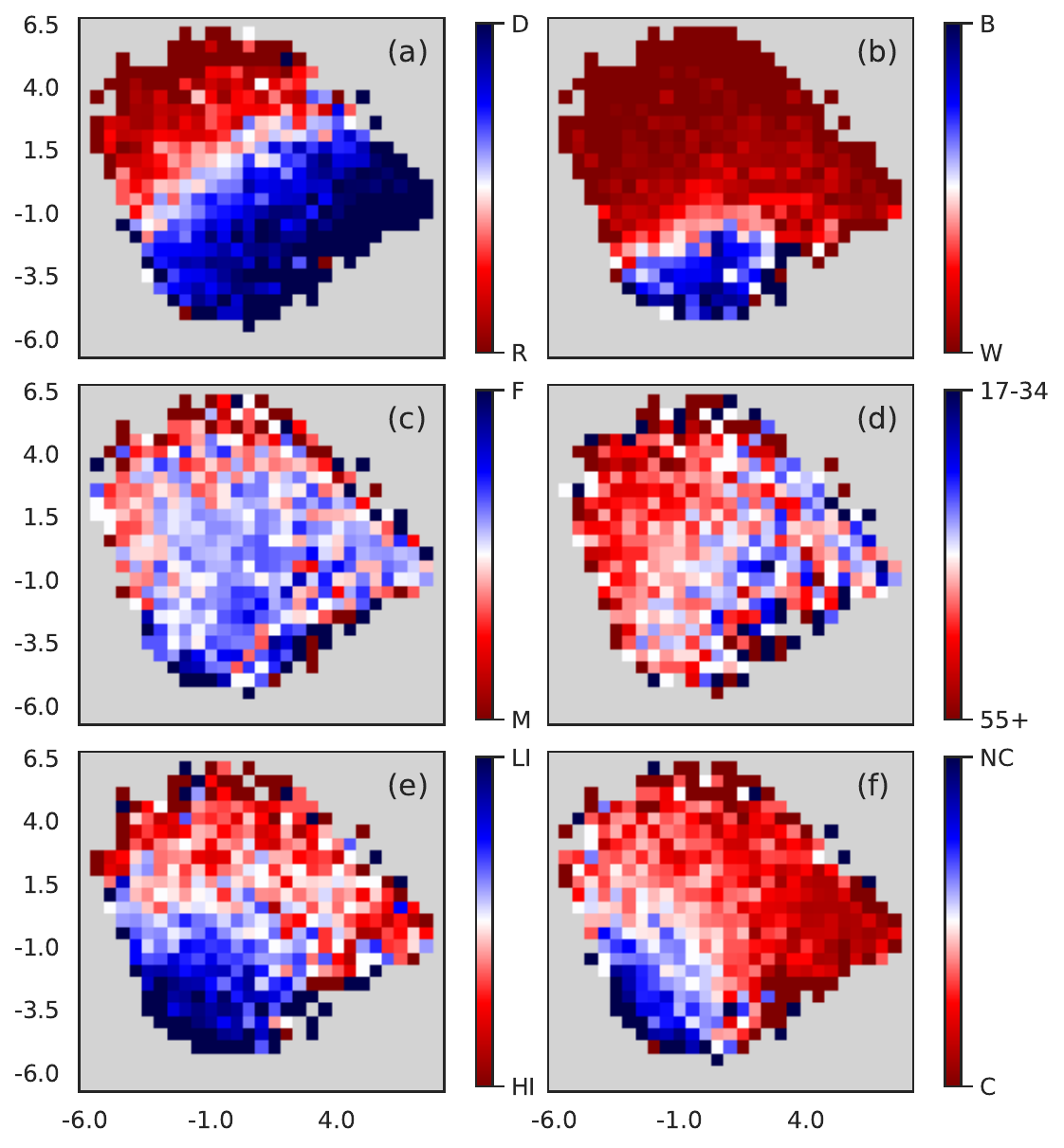}
}
\subfloat[]{
  \includegraphics[width=0.4166\textwidth]{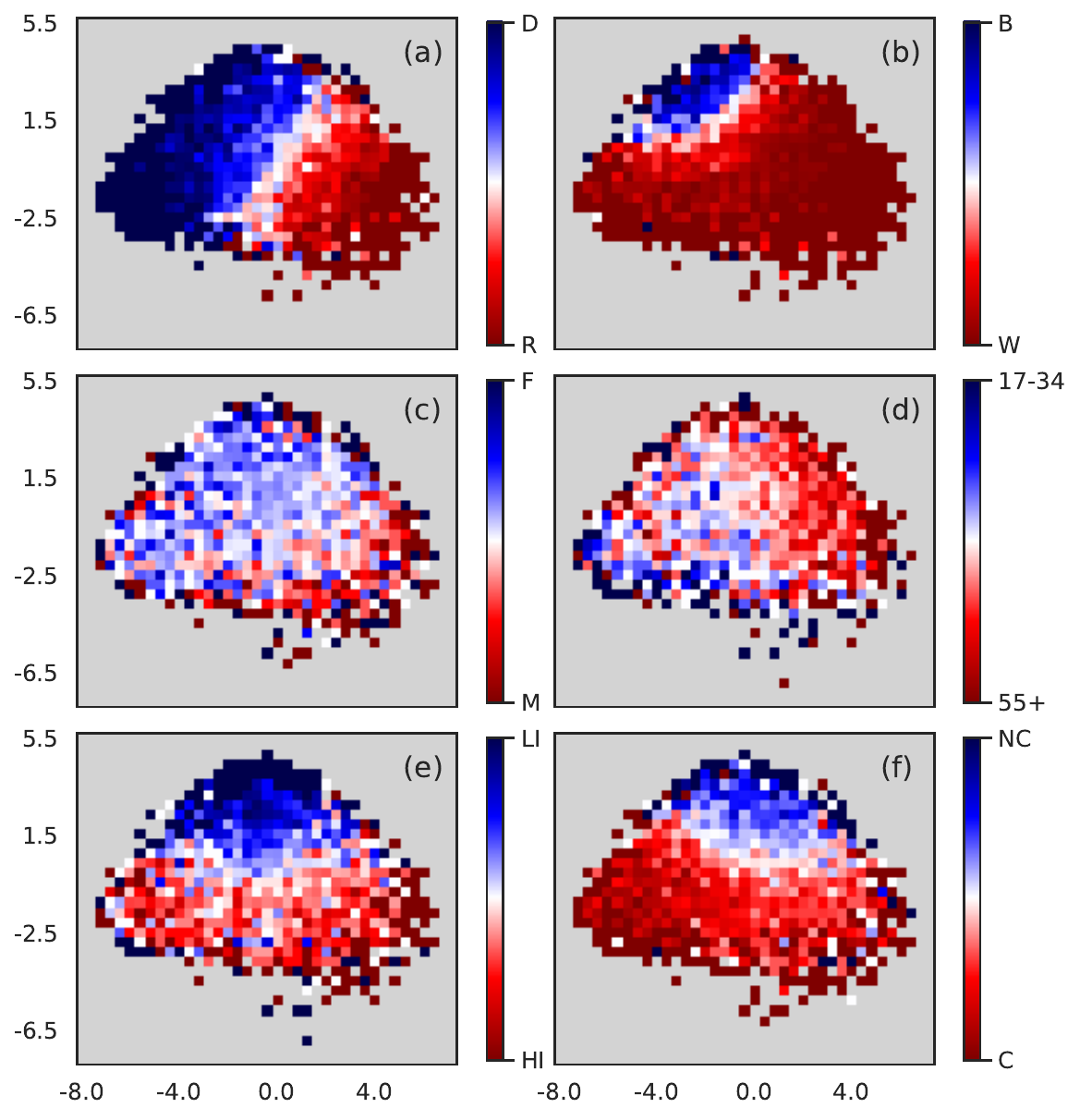}
}

\subfloat[]{
  \includegraphics[width=0.4\textwidth]{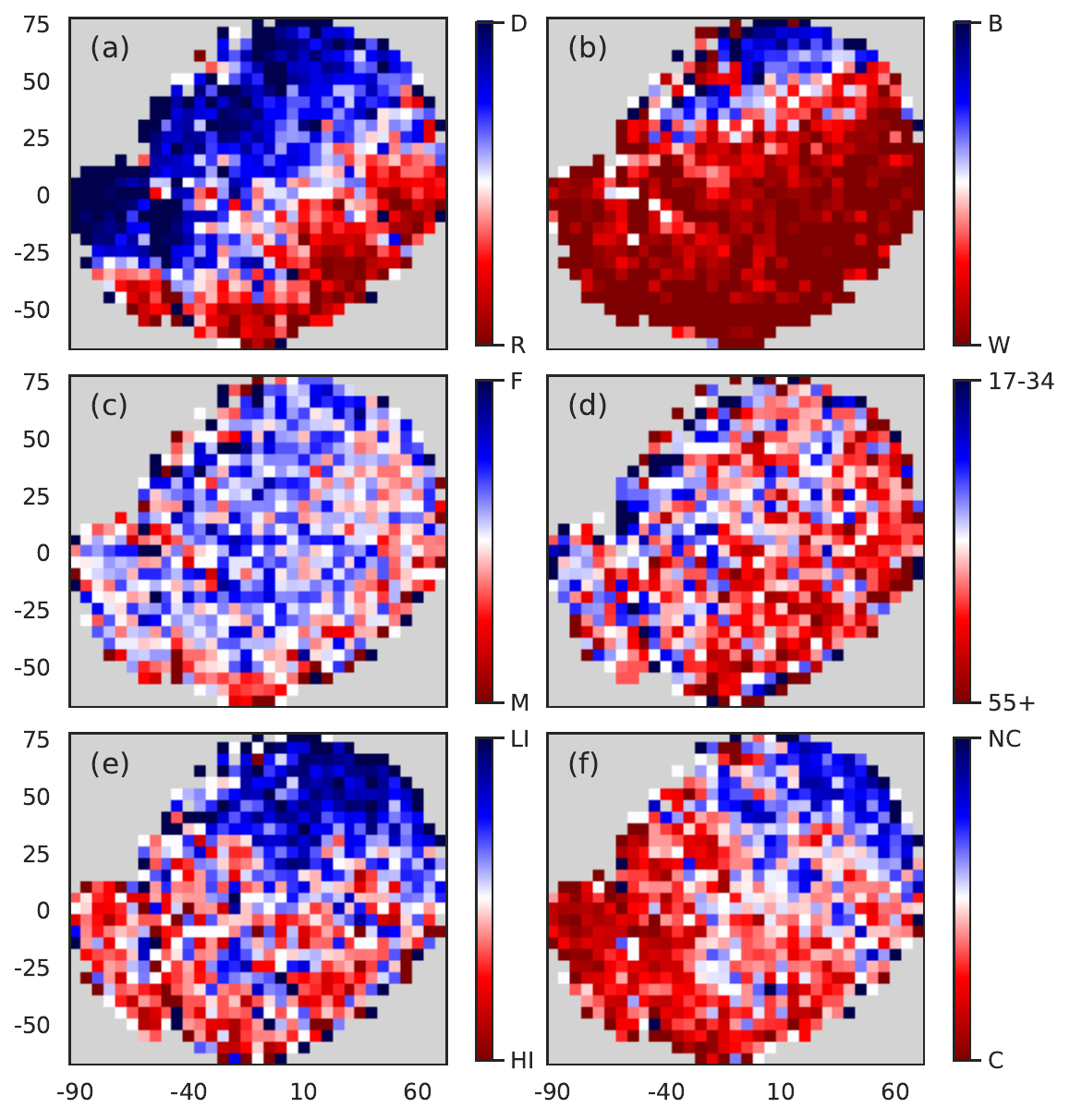}
}
\subfloat[]{
  \includegraphics[width=0.347\textwidth]{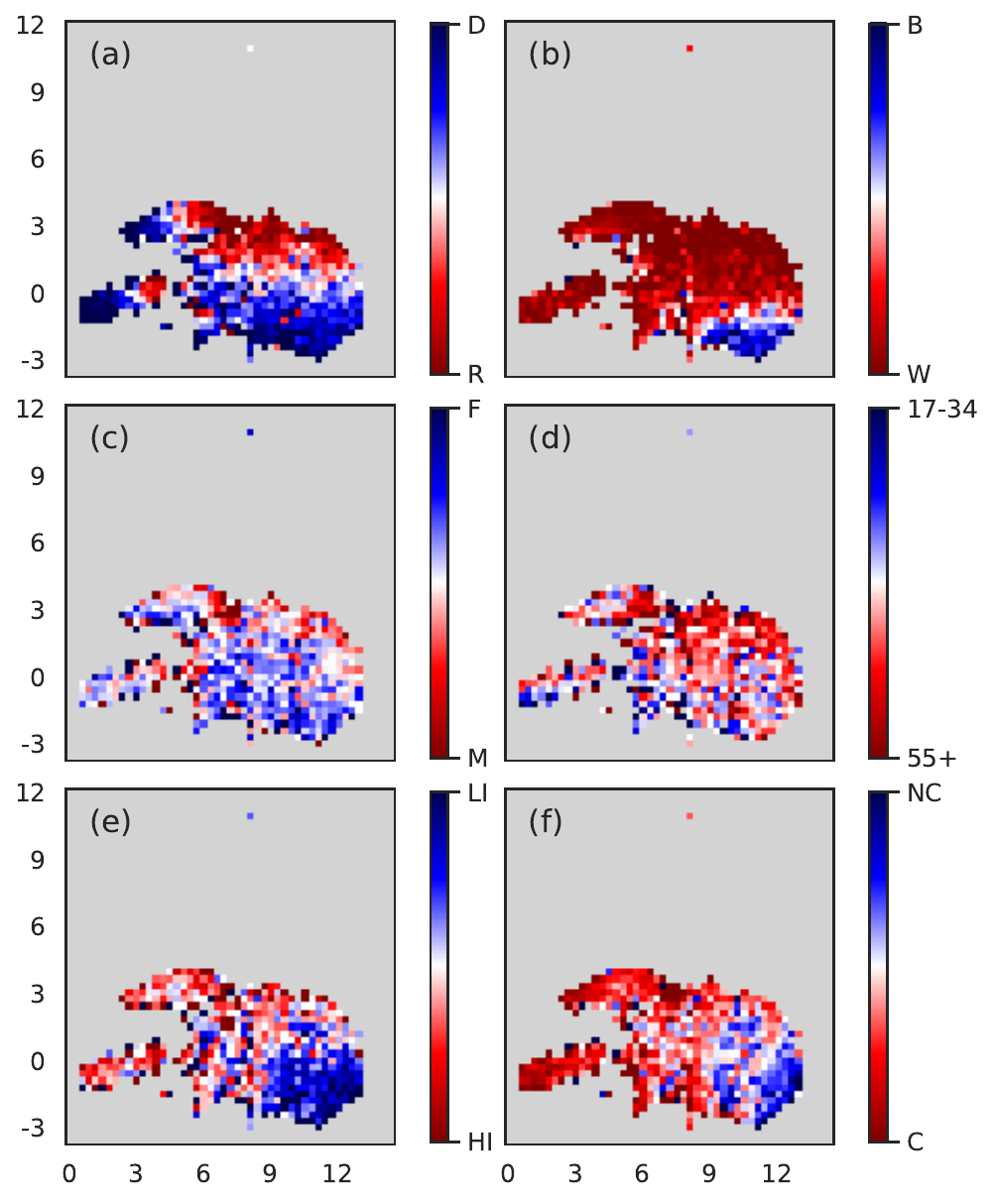}
}
\caption{Density distribution of opposite groups in the ideological space for different attributes, obtained by Isomap with $K = 5$ (i), PCA (ii), t-SNE (iii), and UMAP (iv).}
\label{fig:respondents_density_robust}
\end{figure}

\begin{figure}[h!]
\subfloat[]{
  \includegraphics[width=0.295\textwidth]{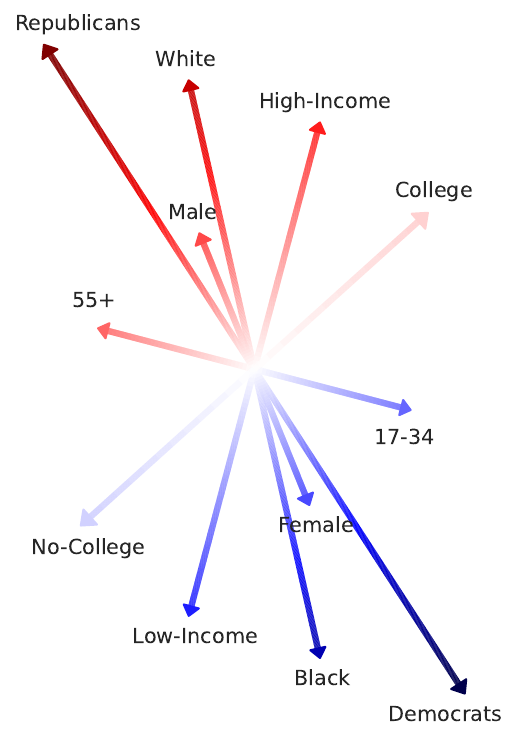}
}
\subfloat[]{
  \includegraphics[width=0.5\textwidth]{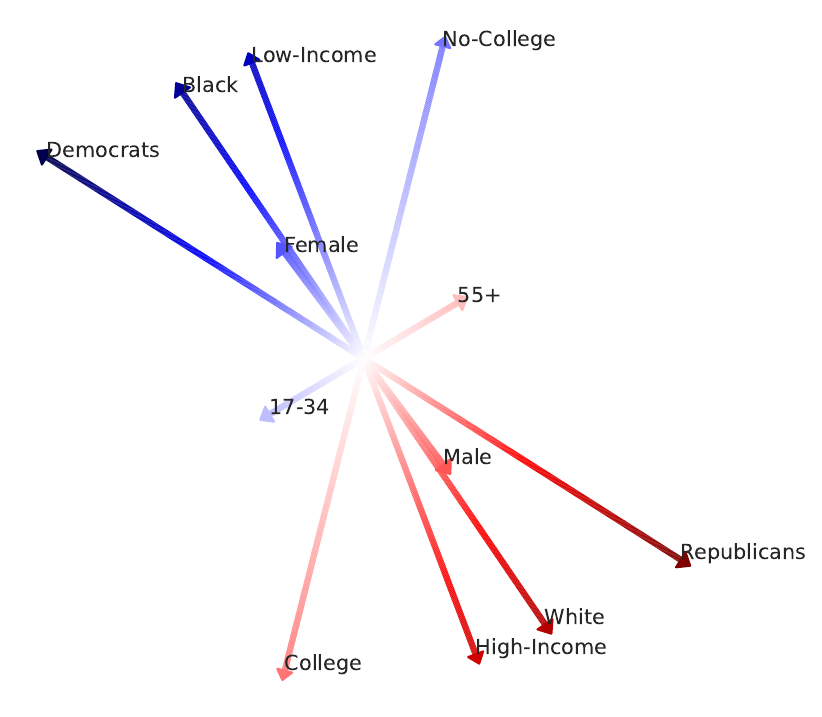}
}

\subfloat[]{
  \includegraphics[width=0.35\textwidth]{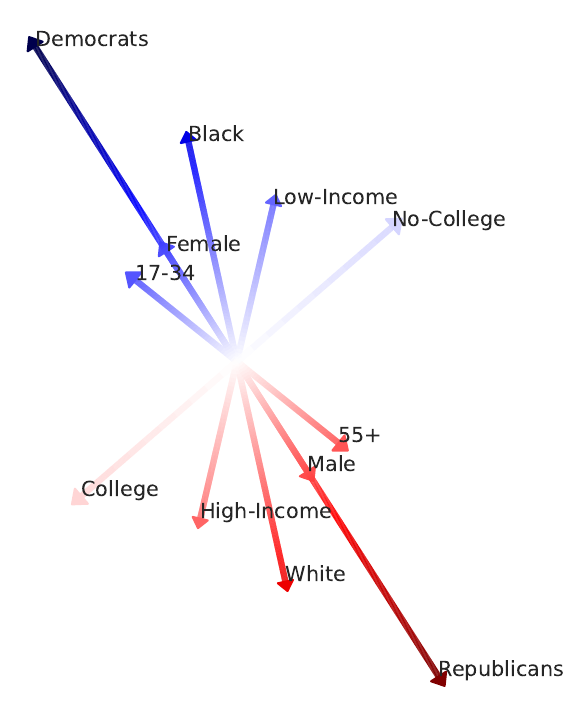}
}
\subfloat[]{
  \includegraphics[width=0.29\textwidth]{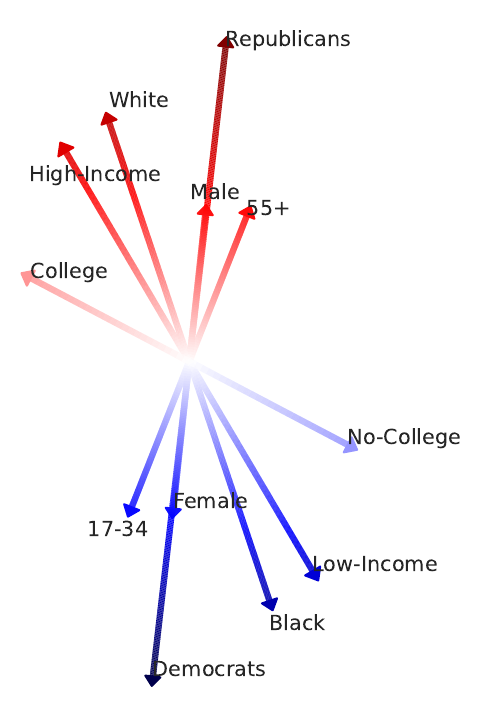}
}
\caption{Gradient vectors for each attribute computed from their opinion distribution obtained by Isomap with $K = 5$ (i), PCA (ii), t-SNE (iii), and UMAP (iv).}
\label{fig:respondents_density_arrows_robust}
\end{figure}

\begin{figure}[h!]
\subfloat[]{
  \includegraphics[width=0.4\textwidth]{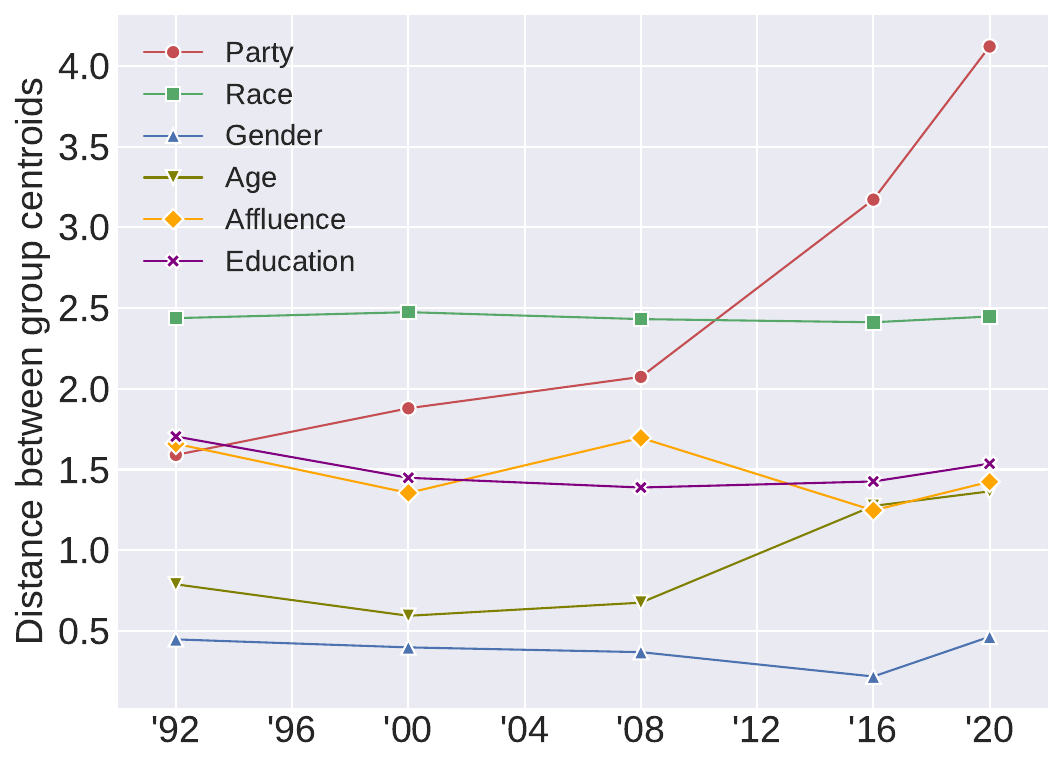}
}
\subfloat[]{
  \includegraphics[width=0.4\textwidth]{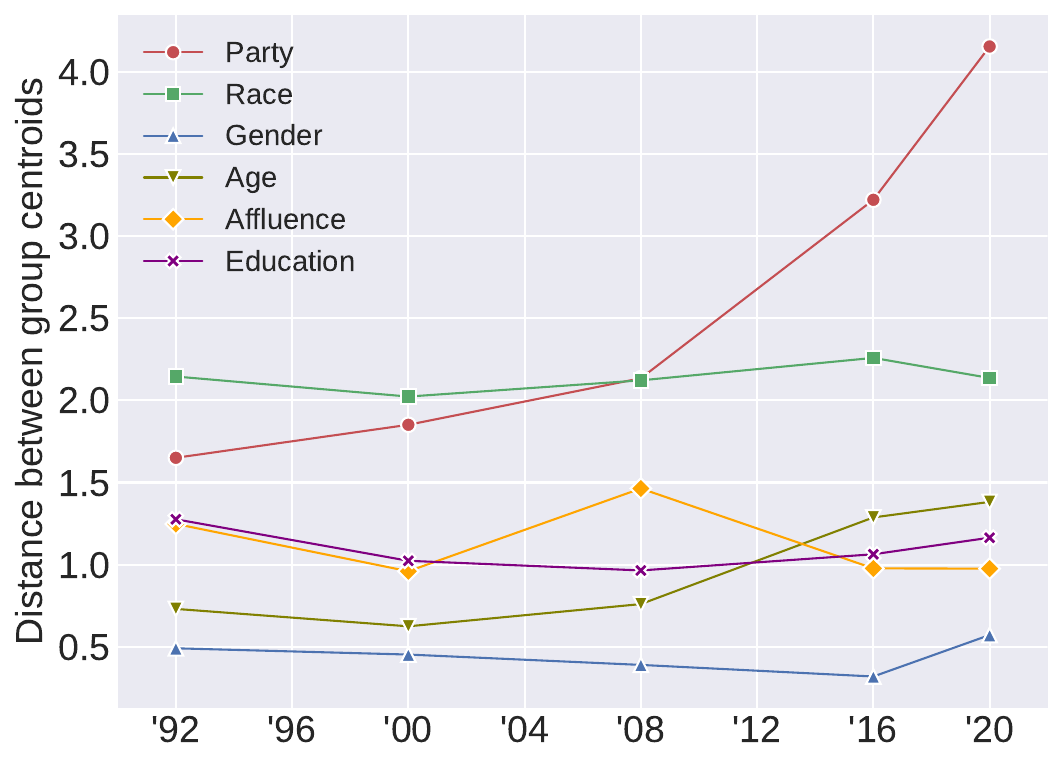}
}

\subfloat[]{
  \includegraphics[width=0.4\textwidth]{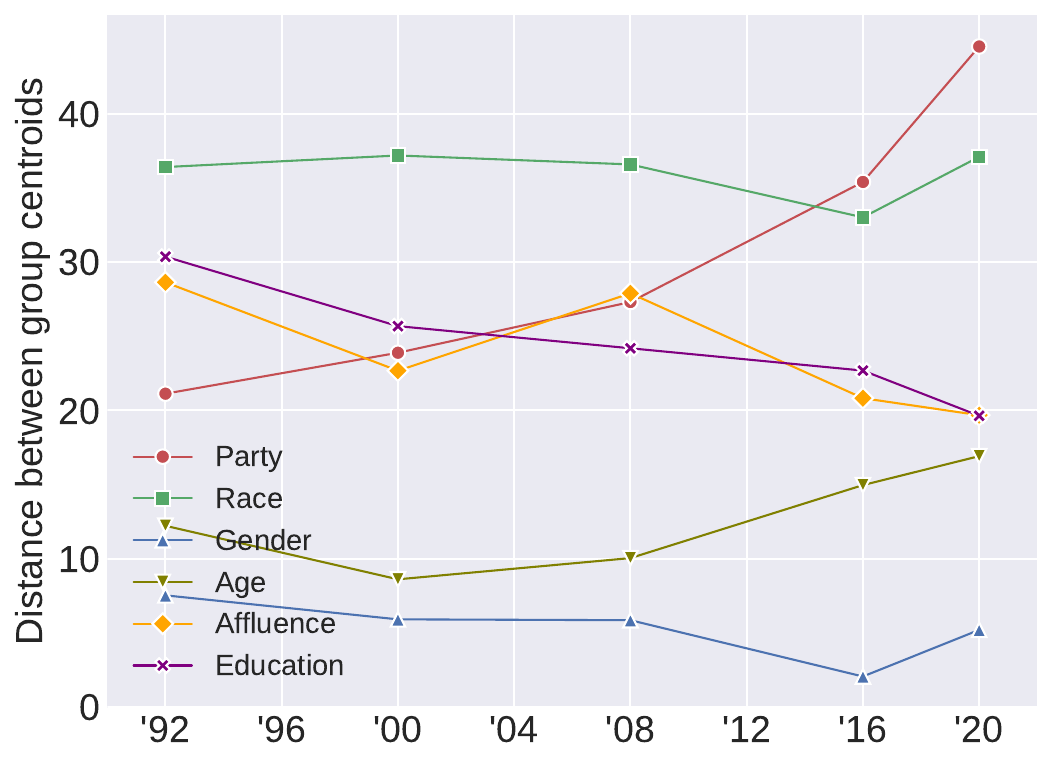}
}
\subfloat[]{
  \includegraphics[width=0.4\textwidth]{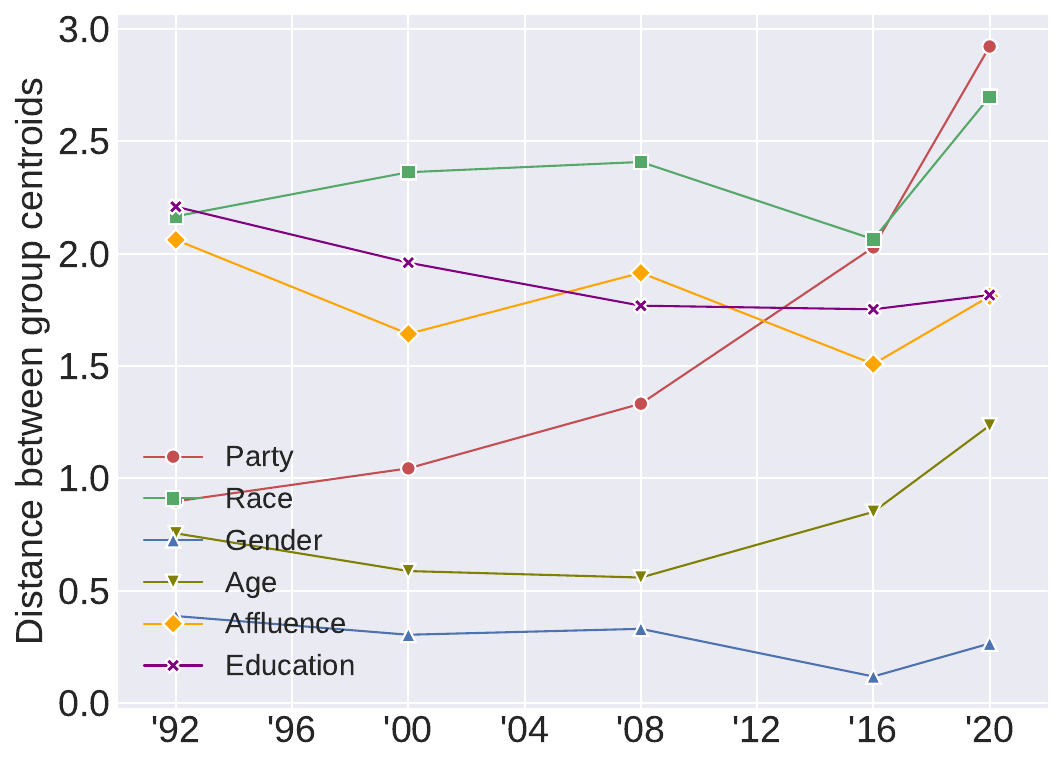}
}
\caption{Euclidean distance between the average opinions (centroids) of the opposite groups of each attribute, for every year, obtained by Isomap with $K = 5$ (i), PCA (ii), t-SNE (iii), and UMAP (iv).}
\label{fig:distances_robust}
\end{figure}

\clearpage

\section{Opinion shift of Democrats}

Supplementary Table~\ref{tab:shiftDemocrats} shows the 5 ANES questions with the largest opinion shift by comparing the answers of Democrats of the years 1992, 2000, 2008 with the ones of 2016, 2020 (see Figure~6 of the main text). 
Since the amount of Democratic respondents was greater in the first time window than in the second one, we have arbitrarily reduced it in order to work with the same number of individuals in both periods, that is 2341 Democrats. 
Taking into account that the answers of respondents range in the numerical interval $\left[ 0,1 \right]$, we calculate the opinion shift before and after 2010 by computing the respective average answers. Therefore, values closer to 0 or 1 represent negative or positive stances, respectively. 
For example, in the case of ``Blacks should not be favored'' with the largest shift, the Democratic respondents in (1992, 2000, 2008) agree more strongly than the ones in (2016, 2020). 
We see that questions are mainly related to minority rights, favoring Blacks and homosexuals, as well as to religion and personal beliefs, moving to a more lay and liberal stance.

\begin{table} [h!]
    \centering
    \begin{tabular}{|p{4cm}|p{4cm}|p{9cm}|}
        \hline
        \hfil \textbf{(1992, 2000, 2008)} & \hfil \textbf{(2016, 2020)} & \textbf{Issue} \\
        \hline
        \hfil 0.656 & \hfil 0.392 & Blacks should not be favored \\
        \hline
        \hfil 0.558 & \hfil 0.301 & Blacks must try harder \\
        \hline
        \hfil 0.630 & \hfil 0.400 & Authority of the bible \\
        \hline
        \hfil 0.681 & \hfil 0.905 & Protection to homosexuals \\
        \hline
        \hfil 0.749 & \hfil 0.534 & Should be more emphasis on traditional values \\
        \hline
    \end{tabular}
    \caption{Opinion shift of Democratic respondents between the first 20 years (1992, 2000, 2008) and the last 10 years (2016, 2020). The 5 ANES questions with the largest variation between both periods are shown.}
    \label{tab:shiftDemocrats}
\end{table}



\end{document}